\documentclass[11pt,letterpaper]{article}
\pdfoutput=1
\usepackage{jcappub}
\usepackage{bm}
\usepackage{amsmath,amssymb}
\usepackage{subfig}
\usepackage{caption}

\newcommand{\xvec}{{\bm{x}}}
\newcommand{\xhat}{\hat{\xvec}}

\newcommand{\kvec}{{\bm{k}}}

\newcommand{\Beq}{\begin{equation}\begin{aligned}}
\newcommand{\Eeq}{\end{aligned}\end{equation}}

\usepackage{graphicx, color}
\usepackage{dcolumn}
\usepackage{bm}
\usepackage[mathlines]{lineno}
\usepackage{amsmath}
\usepackage{amssymb}
\usepackage[numbers,sort&compress]{natbib}
\usepackage{bm}
\usepackage{float}
\usepackage{mathrsfs}
\usepackage{lipsum}
\usepackage{hyperref}
\usepackage[normalem]{ulem}
\hypersetup{
    colorlinks=true,       
    linkcolor=rp,          
    citecolor=green2,        
    filecolor=magenta,      
    urlcolor=green2           
}

\usepackage[all]{hypcap} 
\usepackage{cleveref}

\newcommand{\beq}{\begin{align}}
\newcommand{\eeq}{\end\begin{align}}

\def\lap{\lower.5ex\hbox{$\; \buildrel < \over \sim \;$}}
\def\gap{\lower.5ex\hbox{$\; \buildrel > \over \sim \;$}}

\newcommand{\eref}[1]{eq.~(\ref{#1})}
\newcommand{\erefs}[2]{eqs.~(\ref{#1})~and~(\ref{#2})}
\newcommand{\fref}[1]{figure~\ref{#1}}
\newcommand{\sref}[1]{section~\ref{#1}}
\newcommand{\aref}[1]{appendix~\ref{#1}}

\newcommand{\rref}[1]{ref.~\cite{#1}}

\definecolor{rp}{cmyk}{0.2, 1, 0.6, 0}
\definecolor{rp}{cmyk}{0.2, 1, 0.6, 0}
\definecolor{green2}{cmyk}{0.27, 0, 1, 0.52}

\newcommand{\mpl}{m_{\mathrm{pl}}}

\usepackage{float}
\usepackage{cancel}
\usepackage{soul}
\usepackage{cases}
\usepackage{dsfont}
\usepackage{graphicx}
\usepackage{bm}
\usepackage{color}
\usepackage{amsmath}
\usepackage{adjustbox}
\usepackage{amsmath}
\usepackage{amssymb}
\usepackage{mathrsfs}
\usepackage{textcase}
\usepackage{colortbl}
\usepackage{ulem}
\usepackage{mathrsfs}     
\newcommand{\beqs}{\begin{eqnarray}}
\newcommand{\eeqs}{\end{eqnarray}}
\renewcommand{\l}{\ell}
\newcommand{\m}{m}

\newcommand{\g}{g^2}
\newcommand{\Xbar}{\bar{X}}

\newcommand{\Mvec}{{\bf M}}
\newcommand{\Nvec}{{\bf N}}

\newcommand{\Yvec}{{\bf Y}}
\newcommand{\Psivec}{{\bf \Psi}}
\newcommand{\Phivec}{{\bf \Phi}}
\newcommand{\Avec}{{\bm A}}
\newcommand{\Bvec}{{\bm B}}
\newcommand{\Evec}{{\bm E}}

\newcommand{\Xvec}{{\bm X}}
\newcommand{\evec}{{\bm \epsilon}}

\newcommand{\dvec}{{\bm \nabla}}

\newcommand{\Ocal}{\mathcal{O}}

\newcommand{\Lscr}{\mathscr{L}}

\newcommand{\ba}[1]{\begin{align} #1 \end{align}}
\newcommand{\bes}[1]{\begin{equation}\begin{split} #1 \end{split}\end{equation}}
\newcommand{\bsa}[2]{\begin{subequations}\label{#1}\begin{align} #2 \end{align}\end{subequations}}

\newcommand{\Eref}[1]{Eq.~(\ref{#1})}

\newcommand{\Aref}[1]{Appendix~\ref{#1}}

\title{\huge Photons from dark photon solitons\\ via parametric resonance}

\author{Mustafa A. Amin, }
\emailAdd{mustafa.a.amin@rice.edu}
\author{Andrew J. Long, }
\emailAdd{andrewjlong@rice.edu}
\author{and  Enrico D. Schiappacasse}
\emailAdd{enrico.d.schiappacasse@rice.edu}

\affiliation{Department of Physics and Astronomy, Rice University, Houston, Texas 77005, U.S.A.}

\date{\today}

\abstract{Wave-like dark matter made of spin-1 particles (dark photons) is expected to form ground state clumps called ``vector solitons", which can have different polarizations. In this work, we consider the interaction of dark photons  with photons, expressed as dimension-6 operators, and study the electromagnetic radiation that arises from an isolated vector soliton due to parametric resonant amplification of the ambient electromagnetic field. We characterize the directional dependence and polarization of the outgoing radiation, which depends on the operator as well as the polarization state of the underlying vector soliton.  We discuss the implications of this radiation for the stability of solitons and as a possible channel for detecting mergers of vector solitons through astrophysical observations.\\}

\begin{document}

\maketitle

\tableofcontents

\section{Introduction}

Astrophysical and cosmological observations provide strong evidence for the existence of dark matter~\cite{Bertone:2004pz,Buckley:2017ijx}. However, we do not as yet know the mass, charge, and spin of the constituent dark matter particles. What do astrophysical observations tell us about such properties, especially spin? The electric charge of dark matter cannot be too large \cite{Workman:2022ynf}, whereas the mass cannot be lighter than $O(10^{-19}-10^{-18}\,\rm eV) \,{\rm eV}$~\cite{Dalal:2022a,Amin:2022nlh}. While we do not know the spin of dark matter, an important piece of information connecting the spin and mass of dark matter is  known:  if dark matter is sufficiently light, it cannot be fermionic  since the required occupation number in phase space would be too large \cite{Tremaine:1979}. For bosons, however, light masses are allowed. In the regime when the dark matter mass is sufficiently light ($m\ll \rm eV$), the occupation number of the field in astrophysical settings becomes so large that dark matter is adequately described by a classical, non-relativistic field. Classical, wave dynamical effects become relevant in such settings. Can such wave-effects then be used to infer the spin of bosonic dark matter?

The past decade has seen a  resurgence of effort in exploring wave dynamical effects in non-relativistic, spin-$0$ (i.e. scalar) dark matter. See refs.~\cite{Ferreira:2020fam,Hui:2021tkt} for recent reviews, and  \cite{Schive:2014dra,Du:2016zcv,Mocz:2019pyf,  Niemeyer:2019aqm, Xiao:2021nkb, May:2021wwp} for examples of numerical simulations in a structure formation context. In the case of vector (spin-$1$ or dark photon) dark matter \cite{Co:2018lka, Fabbrichesi:2020wbt,Caputo:2021eaa} a similar numerical exploration is still in its nascent stage~\cite{Amin:2022pzv,Gorghetto:2022sue}. While in a broad sense, the governing equations and the resulting gravitational clustering and growth of structure in non-relativistic vector dark matter is similar to scalars \cite{Adshead:2021kvl, Jain:2021pnk}, the additional number of components in higher spin dark matter ($2s+1$ for a spin-$s$ field) can lead to observationally relevant differences. A larger number of components leads to reduced wave interference, which reduces the variance of density fluctuations in dark matter \cite{Amin:2022pzv}. Such fluctuations can, for example, be probed by dynamical heating of stars \cite{Church:2018sro,Dalal:2022a}. Such effects; however, can also be mimicked to an extent by $n=2s+1$ scalar fields with similar masses \cite{Gosenca:2023yjc}. Furthermore, initial conditions in the early universe do rely on the intrinsic nature (including spin) of the field \cite{Nelson:2011sf,Graham:2015rva,Agrawal:2018vin,Dror:2018pdh,Bastero-Gil:2018uel,Ahmed:2019mjo,Long:2019lwl,Nakayama:2019rhg,Ahmed:2020fhc,Kolb:2020fwh,Adshead:2023qiw}, however, the intrinsic spin (as a spatial vector) is not directly accessible to Newtonian gravity relevant for dark matter in the contemporary universe when it is characteristically non-relativistic. 

To access spin more directly, one must include non-gravitational interactions within the field and/or introduce interactions with other Standard Model fields (or include relativistic corrections). All such effects are typically expected to be small in the case of dark matter.  Nevertheless, the effects of such non-gravitational interactions, even if weak, can be enhanced by the large occupation numbers, densities and coherence length of the dark matter field. These conditions are possible in solitons — coherent field configurations that are long-lived, spatially localized and whose central amplitudes can be much larger than the background density (since the amplitudes do not decay with expansion). For a detailed recent discussion of non-relativistic scalar solitons, see for example \rref{Chavanis:2022fvh} and references therein.

Such solitons have been shown to readily form in light scalar field dark matter via gravitational interactions alone \cite{Schive:2014dra,Levkov:2016rkk}, and recently, also in vector dark matter from cosmological and astrophysical initial conditions \cite{Amin:2022nlh,Gorghetto:2022sue}. Unlike scalar solitons, solitons in vector fields have a richer structure due to the vector nature of the field \cite{Brito:2015pxa,Adshead:2021kvl,Jain:2021pnk}. They can be polarized \cite{Adshead:2021kvl,Jain:2021pnk}, with no particular preference for the polarization in the case of purely gravitational interactions. Such vector solitons typically carry macroscopic amounts of intrinsic spin \cite{Jain:2021pnk}. Non-gravitational self-interactions can lead to preference for one polarization over another, and have been explored in refs.~\cite{Zhang:2021xxa,Jain:2022agt, Wang:2020zur, Jain:2022kwq}. This richness in structure arising from the vector nature of the field provides hope that interactions with Standard Model fields in environments with solitons might lead to interesting, and potentially large spin-dependent effects.

With these considerations in mind, we consider the direct coupling of spin-$1$ dark matter to photons, and explore their implications in an astrophysical environment where solitons are present. We show that such interactions, while very weak, can still lead to resonant production of photons when certain conditions are met. This aspect is similar to the case of resonant photon production from axion stars and miniclusters \cite{Tkachev:2014dpa,Hertzberg:2018zte,Levkov:2020txo, Du:2023jxh}. However, in our case, the polarization pattern of the radiation carries information about the underlying polarization state of the solitons as well  as the specific nature of the interaction. With this preliminary investigation, we elucidate  characteristic features of the electromagnetic radiation (frequency, polarization, spatial patterns of radiation etc.), and the conditions under which such signals are produced. If detected, such signals could provide insight into the underlying spin of dark matter. 

We study resonant photon production from dark photon (i.e. vector) solitons via a variety of dimension-$6$ operators that couple photons and dark photons, within  the framework of effective field theory.  We focus on dimension-$6$ operators since we find that such interactions lead to significant photon production from solitons even in vacuum. Astrophysical implications of a more natural dimension-$4$ operator: gauge kinetic mixing \cite{Holdom:1985ag,Foot:1991kb}, has been explored extensively in the literature (e.g.~\cite{Dubovsky:2015cca,Kovetz:2018zes,Bhoonah:2018gjb,Wadekar:2019mpc}), albeit in non-solitonic settings. Photon production from such a coupling is also of interest in the presence of solitons, and might lead to enhanced signals. 
Furthermore, our effort here is complementary to the significant ongoing effort to detect light dark photon dark matter in terrestrial settings \cite{Antypas:2022asj}.

The remainder of the article is organized as follows.  
The content of \sref{sec:model} establishes the scope of the problem:  we specify the model for massive dark photons interacting with electromagnetism, we discuss a possible ultraviolet embedding for the dimension-6 operators that we study, and we present the spatially-localized polarized vector soliton configurations. 
The core results of our study are presented in \sref{Sec:presonance}, which includes our analysis of the electromagnetic field's equation of motion using Floquet theory and our predictions for the Floquet exponents arising from parametric resonance of a dark photon homogeneous configuration with either linear or circular polarization. 
In \sref{sec:vector_soliton}, we apply previous results to study electromagnetic radiation from polarized vector solitons and discuss the possible astrophysical signatures. 
In \sref{sec:Conclusion}, we conclude and summarize key points of our work.  
\Aref{app:details} contains details of the homogeneous Floquet analysis, \aref{app:spherical} contains the modified Floquet analysis for an inhomogeneous vector soliton, and \aref{app:fuzzy} includes an extension of our work to the case of fuzzy dark photon dark matter.

\section{Modeling dark photon interactions with light}
\label{sec:model}

We are interested in the interactions of a massive spin-1 dark photon with electromagnetism.  
Consider a massive real vector field $X_\mu(x)$, which we call the dark photon field.
The properties and interactions of these particles are encoded in the action 
\ba{
\label{eq:action}
    S[X_\mu(x), A_\mu(x), \text{g}_{\mu\nu}(x)] = \int \! \mathrm{d}^4 x \, \sqrt{-\text{g}} \biggl[ - \frac{1}{4} X_{\mu\nu} X^{\mu\nu} - \frac{1}{2} m^2 X_\mu X^\mu - \frac{1}{4} F_{\mu\nu} F^{\mu\nu} + \frac{1}{2} \mpl^2 \text{R} + \Lscr_\mathrm{int} \biggr]}
where $X_{\mu\nu} = \nabla_\mu X_\nu - \nabla_\nu X_\mu$ is the dark photon field strength tensor, $F_{\mu\nu} = \nabla_\mu A_\nu - \nabla_\nu A_\mu$ is the electromagnetic field strength tensor, $\text{R}$ is the Ricci scalar, and indices are raised and lowered with the metric $\text{g}_{\mu\nu}(x)$.    
We work in natural units where $\hbar=c=1$ are set to one, $\mpl=1/\sqrt{8\pi G_N}$ is the reduced Planck mass, and (-\,+\,+\,+) is the metric signature. 
 We also write $X_\mu = (X_0, \Xvec)$ and $\partial_\mu f = (\dot{f}, \dvec f)$.  
We consider small values of the mass parameter $m \ll 10 \ \mathrm{eV}$ corresponding to light dark photons.  Extending earlier work on dark photons, we allow for interactions between $X_\mu(x)$ and the electromagnetic field $A_\mu(x)$, which is represented by $\Lscr_\mathrm{int}$.  
We enumerate the relevant interaction operators in Sec.~\ref{sub:interactions}; these include $\Lscr_\mathrm{int} \supset F_{\mu\nu} F_{\rho\sigma} X_\alpha X_\beta$ and $F_{\mu\nu} F_{\rho\sigma} \partial_\alpha X_\beta$ where the Lorentz indices may be contracted with various combinations of the inverse metric and Levi-Civita symbol.  

\subsection{Non-relativistic modes of the dark photon field}
\label{sub:nr_modes}

We are interested in the dark photon as a candidate for the cold dark matter. 
In the systems of interest, only non-relativistic modes of the dark photon field will propagate; these modes have small wavenumbers $k \ll m$ and large de Broglie wavelengths $\lambda \gg 2\pi / m$. 
This observation motivates a perturbative expansion in powers of the dark photon field's spatial gradient; the parametric relations are $|\dvec X_\mu| \sim \lambda^{-1} X_\mu \ll m X_\mu \sim \dot{X}_\mu$.  
We work to leading order in this expansion, which effectively amounts to setting $\dvec X_\mu = 0$.\footnote{We work in the zero spatial gradient approximation locally, but indirectly take spatial gradients into account by including the finite size effects of dark photon configurations in the phenomenology.}  
The temporal component of the dark photon field, $X_0(x)$, is non-dynamical in the theories that we study. In the time component of the Euler-Lagrange equations, $\ddot{X}_0(x)$ cancels out and  
its equation of motion is an algebraic constraint equation, which has the solution $X_0 = \bigl( \nabla^2 - m^2 \bigr)^{-1} \bigl( \dvec \cdot \dot{\Xvec})$, neglecting gravitational and electromagnetic interactions.  
Working to leading order in the gradient expansion, we set $X_0(x) = 0$.  

\subsection{Interactions with electromagnetism}
\label{sub:interactions}

Since we seek to study electromagnetic radiation from vector solitons, it is necessary to introduce a coupling between the dark photon field $X_\mu(x)$ and the electromagnetic field $A_\mu(x)$.  
Working in the context of effective field theory (EFT), we consider all operators that are consistent with electromagnetic gauge invariance, and we organize the operators based on their mass dimension.  
The only such operator with mass dimension-4 is the so-called gauge-kinetic mixing~\cite{Holdom:1985ag,Foot:1991kb}
\ba{
    \Lscr_\mathrm{int}^{(4)} 
    & \supset 
    F_{\mu\nu} X_{\alpha\beta} 
    \;,
}
where $F_{\mu\nu} = \partial_\mu A_\nu - \partial_\nu A_\mu$ is the usual electromagnetic field strength tensor and $X_{\alpha\beta} = \partial_\alpha X_\beta - \partial_\beta X_\alpha$.
The Lorentz indices can be contracted using any combination of the diagonal inverse Minkowski metric $\eta^{\mu\nu}$ and the totally-antisymmetric Levi-Civita symbol $\epsilon^{\mu\nu\rho\sigma}$; we normalize $-\eta^{00} = \eta^{11} = \eta^{22} = \eta^{33} = \epsilon^{0123} = 1$.  
The gauge kinetic mixing can be exchanged for a coupling to charged matter by performing a field redefinition.  
In this work we consider systems in the absence of free charges, and the gauge-kinetic mixing operators do not lead to electromagnetic radiation from a dark photon field.
At mass dimension-5 there are no operators coupling the vector soliton to electromagnetism, since such operators would carry an odd number of Lorentz indices, which cannot be fully contracted using only the two-index metric and the four-index Levi-Civita symbol.  
At dimension-6 the following operators are available:
\bes{
    \Lscr_\mathrm{int}^{(6)} 
    & \supset 
    F_{\mu\nu} F_{\rho\sigma} X_\alpha X_\beta
    \ , \quad
    F_{\mu\nu} F_{\rho\sigma} \partial_\alpha X_\beta
    \ , \quad
    F_{\mu\nu} X_\rho X_\sigma \partial_\alpha X_\beta
    \ , \quad
    F_{\mu\nu} \partial_\rho X_\sigma \partial_\alpha X_\beta
    \ , \quad
    F_{\mu\nu} \partial_\rho \partial_\sigma \partial_\alpha X_\beta\,.
}
The third, fourth, and fifth operators involve only one factor of the electromagnetic field $A_\mu(x)$.    
In the presence of a background dark photon field $X_\mu(x)$, these operators provide a source for $A_\mu(x)$.  
The radiation arising from such source terms is highly suppressed for long-wavelength background fields if plasma effects can be neglected~\cite{Amin:2021tnq}, and we do not discuss these operators further here. 
  
The dimension-6 operators that we study are summarized as follows:\footnote{Some of these operators are related to one another using integration by parts (dropping total derivatives) and equations of motion.  For the non-relativistic dark photon field, a few other operators reduce to one of these; for instance $F_{\mu\rho} \tilde{F}^{\nu\rho} X^\mu X_\nu \approx -\Ocal_1$.  }  
\bsa{eq:operators}{
    \Ocal_1 
    & = 
    -\tfrac{1}{2} F_{\mu\nu} \tilde{F}^{\mu\nu} (X \cdot X)
    & \approx & \ 
    2 (\Evec \cdot \Bvec) (\Xvec \cdot \Xvec) \\ 
    \Ocal_2 
    & = 
    -\tfrac{1}{2} F_{\mu\nu} F^{\mu\nu} (X \cdot X)
    & \approx & \ 
    (\Evec \cdot \Evec) (\Xvec \cdot \Xvec) - (\Bvec \cdot \Bvec) (\Xvec \cdot \Xvec) \\ 
    \Ocal_3 
    & = 
    F_{\mu\rho} F^{\nu\rho} X^\mu X_\nu 
    & \approx & \ 
    (\Bvec \cdot \Bvec) (\Xvec \cdot \Xvec) - (\Evec \cdot \Xvec)^2 - (\Bvec \cdot \Xvec)^2 \\ 
   \Ocal_4 
    & = 
    \tilde{F}_{\mu\rho} \tilde{F}^{\nu\rho} X^\mu X_\nu 
    & \approx & \ 
    (\Evec \cdot \Evec) (\Xvec \cdot \Xvec) - (\Evec \cdot \Xvec)^2 - (\Bvec \cdot \Xvec)^2 \\ 
    \Ocal_5 
    & = 
    F_{\mu\rho} F^{\nu\rho} \partial^\mu X_\nu 
    & \approx & \ 
    (\Evec \times \Bvec) \cdot \dot{\Xvec} 
    \;.
}
To move from the Lorentz-covariant expressions to the 3-vector expressions, we have dropped terms containing $X_0$ and spatial gradients $\dvec X_\mu$, which is an excellent approximation for non-relativistic modes of the dark photon field.  

We write $\Lscr_\mathrm{int} = g^2 \Ocal_i$ and we study the effect of each operator one at a time.  
Validity of the effective field theory, which allows us to neglect the effects of dimension-8 (and higher-order) operators, requires the coupling $g^2$ to remain sufficiently small.  
Moreover, we consider systems in which the dark photon field acquires a nonzero vacuum expectation value $\langle \Xvec \rangle \sim \Xbar$, which causes these dimension-6 operators to renormalize lower-order operators; for instance, $\Ocal_2$ modifies the electromagnetic kinetic term.  
To ensure that these modifications are negligible, and that the EFT remains valid, we impose 
\ba{\label{eq:EFT_condit}
    \g \Xbar^2 \ll 1 
    \;,
}
where $\Xbar$ is interpreted as the typical amplitude of the dark photon field $\Xvec(t,\xvec)$.  

\subsection{Ultraviolet embedding}
\label{sub:uv}

Each of the operators in \eref{eq:operators} is used to construct an effective field theory with $\Lscr_\mathrm{int} = \g \Ocal_i$, and we study the resultant electromagnetic radiation from a non-relativistic dark photon field.  
Our analysis is independent of the EFT's ultraviolet (UV) embedding, except insofar as we are justified to `turn on' each operator, one at a time.  
Nevertheless, it is interesting to remark that these operators can arise from a simple, renormalizable theory in the UV.  
In the remainder of this short section, we offer a concrete UV embedding for operator $\Ocal_2$.  

Consider the following theory.  
Suppose that $X_\mu$ is the vector potential associated with a dark $\mathrm{U}(1)_d$ gauge symmetry, and suppose that the UV theory includes a dark Higgs field $\phi(x)$ with $D_\mu \phi = \partial_\mu \phi - i g_d X_\mu \phi$.  
If the dark Higgs acquires a nonzero vacuum expectation value $\langle \phi \rangle = v_d / \sqrt{2}$, then operator $\Ocal_2$ can arise from the dimension-8 operator: 
\ba{
    \Lscr_8 = 
    - \tfrac{1}{8} \, M^{-4} \, \bigl| D_\alpha \phi \bigr|^2 F_{\mu\nu} F^{\mu\nu} 
    \;.
}
The operator coefficients in our EFT are parametrically $\g \sim g_d^2 v^2 / M^4 \sim m^2 / M^4$ where $m \sim g_d v$ is the mass scale of the dark photon and $M$ is the UV scale of new physics.  
The dimension-8 operator, in turn, may arise from a renormalizable theory of charged fermions $\psi$ and $\chi$ with a Yukawa coupling $- y \phi \bar{\psi} \chi + \mathrm{h.c.}$.  
A one-loop box graph generates $\Lscr_8$ upon integrating out the fermions.  
Assuming that the fermions have comparable mass $m_\chi \sim m_\psi$ and electromagnetic charge $q_\psi e$, the box graph is parametrically $M^{-4} \sim y^2 q_\psi^2 e^2 / 16 \pi^2 m_\psi^4$.  
Finally we arrive at a parametric estimate for the operator coefficients in our EFT: $g \sim y q_\psi e m / 4 \pi m_\psi^2$.  

In the next section, we show that operators $\Ocal_1$ through $\Ocal_4$ lead to resonance as long as $g \mpl \gg 1$.  
For a fiducial set of parameters, we estimate $g \mpl \sim (y/1) (q_\psi / 10^{-14}) (m / 10^{-6} \ \mathrm{eV}) (m_\psi / \mathrm{keV})^{-2}$.  
These parameters are chosen to reflect the constraints on millicharged particles, which place tight upper limits on $q_\psi$ across a wide range of $m_\psi$ values~\cite{Vogel:2013raa}.  
The strongest limits from stellar cooling plateau to $q_\psi \lesssim 10^{-14}$ for $m_\psi$ below $10 \ \mathrm{keV}$; lowering $m_\psi$ further does not strengthen the $q_\psi$ limit.  
These estimates imply that a sufficiently large dimensionless coupling $g \mpl \gg 1$ can be achieved if the `UV' embedding includes sufficiently light and weakly-charged fermions.  
Despite the small value of $m_\psi$ compared to the Standard Model particle content, the EFT approach remains valid while the fermion mass is much larger than the dark photon mass, i.e. $m_\psi \sim \mathrm{keV} \gg m \sim \mu\mathrm{eV}$.  

\subsection{Polarized vector solitons}
\label{sub:soliton}

In the nonrelativistic regime, the equations of motion for the dark photon field $\Xvec(t,\xvec)$ and the gravitational field is a Schr\"odinger-Poisson system ~\cite{Adshead:2021kvl,Jain:2021pnk}. 
These equations admit spatially-localized solutions with spherically-symmetric density profiles, which correspond to gravitationally bound and coherent clumps of dark photons that are ground states of the system at fixed particle number~\cite{Adshead:2021kvl}. 
Such solitons have spatially-independent polarization of the field, with linear and circular polarization being the extremal cases. 
These have been called polarized vector solitons, and they typically carry macroscopic amount of spin angular momentum~\cite{Jain:2021pnk}.  
A general polarized vector soliton field configuration takes the form 
\ba{\label{eq:X_ansatz}
    \Xvec(t,\xvec) & = \frac{1}{2} \sum_{a} \Bigl[c^{(a)} \, X(r) \, e^{-i (m-\mu) t} \, \evec^{(a)} + \mathrm{h.c.} \Bigr] 
     \;,
}
where $r = |\xvec|$ is the radial distance from the center of the soliton, the index $a$ labels the three polarization modes, $\evec^{(a)}$ are the corresponding polarization unit vectors that are constants, and $c^{(a)}$ are c-number coefficients that are normalized by $\sum_a |c^{(a)}|^2 = 1$.  
The real and positive parameter $\mu$, called the chemical potential, controls the field amplitude via the radial field profile $X(r)$.  
Note that the field amplitude oscillates in time with an angular frequency $\omega = m - \mu$.  
Validity of the non-relativistic approximation requires 
\ba{
    \mu/m \ll 1 
    \qquad \text{and} \qquad 
    \omega \approx m 
    \;.
}
For instance, vector soliton formation by the collapse of Hubble-scale inhomogeneities at radiation-matter equality~\cite{Gorghetto:2022sue} would give $\mu \sim H_\mathrm{eq} \approx 2\times 10^{-28} \ \mathrm{eV}$, which is far below the fiducial mass scale $m \approx 10^{-6} \ \mathrm{eV}$. 
For solitons forming in nonlinear environments inside dark matter halos, the chemical potential is expected to be comparable to the typical kinetic energy per particle in the environment leading to $\mu/m \sim v^2 \sim 10^{-6}$~\cite{Schive:2014dra,Levkov:2016rkk}. 

The radial field profile $X(r)$ and the non-dynamical Newtonian potential $\Phi(r)$ are required to solve the static Schr\"odinger-Poisson system of equations.
For each polarization mode, a one-parameter family of solutions are labeled by the chemical potential $\mu$, which sets the amplitude of $X(r)$ and thus also $\Xvec(t,\xvec)$.  
These solutions are well-approximated by the empirical fitting formula~\cite{Schive:2014dra,Amin:2022pzv} 
\ba{\label{eq:Psia}
    X(r) \simeq \frac{\Xbar}{(1 + 0.077 \, \mu m r^2)^4}
    \qquad \text{with} \qquad 
    \Xbar \simeq 2.04 \, \mpl \, \Bigl( \frac{\mu}{m} \Bigr) 
    \;.
}
The localized soliton solution has a finite gravitational binding energy $E$, total mass $M$, and full width at half maximum $R$ that are given by~\cite{Jain:2021pnk}\footnote{The numerical factors  are more accurate than those provided in \cite{Jain:2022agt,Amin:2022pzv}.} 
\ba{\label{eq:EMR}
    E \approx -20.8\, \frac{\mpl^2}{m} \Bigl( \frac{\mu}{m} \Bigr)^{3/2} 
    \ , \qquad 
    M \approx 62.3\, \frac{\mpl^2}{m} \Bigl( \frac{\mu}{m} \Bigr)^{1/2}
    \ , \quad \text{and} \qquad 
    R \approx 3.16 \, \frac{1}{m} \Bigl( \frac{\mu}{m} \Bigr)^{-1/2}
    \;.
}
 which have an error of $\lesssim 10\%$.
Since $\mu/m \ll 1$ it follows that $|E| \ll M$, implying that the particles in the vector soliton are cold, and that there are approximately $N \approx M/m$ constituent particles.  
The average binding energy per particle is $E/N \approx -0.33 m (\mu/m)$.  
To ensure that the soliton is a many-particle state, $N \gg 1$, the chemical potential is bounded from below as $\mu/m \gg (m/\mpl)^4$, which is easily satisfied, since $m \lll \mpl$ for the parameters of interest. 

The three polarization unit vectors $\evec^{(a)}(\hat{\xvec})$ form an orthonormal basis.  
Two convenient basis choices are 
\bes{\label{eq:vec}
    \evec^{(x)} = \begin{bmatrix} 1 \\ 0 \\ 0 \end{bmatrix} 
    \! , \ \  
    \evec^{(y)} = \begin{bmatrix} 0 \\ 1 \\ 0 \end{bmatrix} 
    \! , \ \  
    \evec^{(z)} = \begin{bmatrix} 0 \\ 0 \\ 1 \end{bmatrix} 
    \qquad \text{and} \qquad 
    \evec^{(-)} = \frac{1}{\sqrt{2}} \begin{bmatrix} 1 \\ -i \\ 0 \end{bmatrix} 
    \! , \ \  
    \evec^{(0)} = \begin{bmatrix} 0 \\ 0 \\ 1 \end{bmatrix} 
    \! , \ \  
    \evec^{(+)} = \frac{1}{\sqrt{2}} \begin{bmatrix} 1 \\ i \\ 0 \end{bmatrix} 
    \;.
}
They correspond to linear polarization along each of the three coordinate axes and circular polarization with respect to the third $z$ axis.  
If non-gravitational interactions can be neglected, each of these six modes is degenerate~\cite{Jain:2021pnk}. 
We do not consider the `hedgehog' configuration $\evec = \hat{\xvec}$~\cite{Brito:2015pxa}, since it corresponds to a state of higher energy.  
For example, using the circular polarization basis allows the polarized vector soliton field configuration to be written as
\ba{\label{eq:W2}
    \Xvec(t,r) = X(r) \left( 
    \frac{|c^{(-)}|}{\sqrt{2}} \begin{bmatrix}
    \cos(\omega t - \mathrm{arg} \, c^{(-)}) \\
    -\sin(\omega t - \mathrm{arg} \, c^{(-)})\\
    0
    \end{bmatrix} 
    + |c^{(0)}| \begin{bmatrix}
    0 \\
    0 \\
    \cos(\omega t - \mathrm{arg} \, c^{(0)})
    \end{bmatrix} 
    + \frac{|c^{(+)}|}{\sqrt{2}} \begin{bmatrix}
    \cos(\omega t - \mathrm{arg} \, c^{(+)}) \\
    \sin(\omega t - \mathrm{arg} \, c^{(+)})\\
    0
    \end{bmatrix}
    \right) 
    \;.
}
where $c^{(a)} = |c^{(a)}| e^{i \, \mathrm{arg} \, c^{(a)}}$ and $\omega = m-\mu$.  

\section{Electromagnetic radiation via parametric resonance}
\label{Sec:presonance}

Interactions between the dark photon field and the electromagnetic field allow for electromagnetic radiation to arise from a dynamical dark photon field configuration, even in the absence of charged matter.  
We are concerned with the operators appearing in \eref{eq:operators}.  
In the background of the oscillating dark photon field $\Xvec(t,\xvec)$, these operators induce a time-dependent equation of motion for the electromagnetic field. 
This leads to the phenomenon of parametric resonance, which can be studied using Floquet theory.  
Fourier modes of the electromagnetic field that fall into resonance bands experience an exponential amplification $\Avec_\kvec(t) \propto e^{\mu_{\kvec} t}$, where $\mu_{\kvec}$ are the Floquet exponents, allowing a weak seed field to be transformed into electromagnetic radiation. 
This radiation extracts energy from the dark photon field, which impacts its lifetime while also providing a signal that would make dark photon evaporation possibly detectable from Earth.  

In the remainder of this section, we apply known techniques from Floquet theory to develop an analytical formalism that allows us to study parametric resonance of the electromagnetic field coupled to a dark photon field.  
We derive expressions for the Floquet exponents $\mu_{\kvec}$ assuming different polarization configurations for the dark photon field.  
As a simplifying approximation, throughout this section we treat the dark photon field as spatially homogeneous: 
$\Xvec(t,\xvec) = \Xvec(t)$.  
In the following sections, we discuss how our results should be adapted for the study of inhomogeneous polarized vector solitons.  

\subsection{Electromagnetic equation of motion}
\label{sub:EOM}

For each of the five operators that we study, the electromagnetic field's equation of motion is linear.  
Working in the Coulomb gauge $\dvec \cdot \Avec = 0$, the equation of motion admits a Fourier representation: 
\ba{\label{eq:OPQ_equation}
    \mathbb{O}_{ij} \ddot{A}_j + \mathbb{P}_{ij} \dot{A}_j + \mathbb{Q}_{ij} A_j = 0 \;,
}
where the matrix coefficients are 
\bsa{eq:OPQ_list}{
    \mathbb{O}_{ij} & = \begin{cases}
    \delta_{ij} & \text{, \ $\Lscr_\mathrm{int} = \g \Ocal_1$} \\ 
    \delta_{ij} + \bigl( 2 \g |\Xvec|^2 \bigr) \, \delta_{ij} & \text{, \ $\Lscr_\mathrm{int} = \g \Ocal_2$} \\ 
    \delta_{ij} + \bigl( - 2 \g \bigr) \, X_i X_j + \bigl( 2 \g \frac{\kvec \cdot \Xvec}{|\kvec|^2} \bigr) \, k_i X_j & \text{, \ $\Lscr_\mathrm{int} = \g \Ocal_3$} \\ 
    \delta_{ij} + \bigl( 2 \g |\Xvec|^2 \bigr) \, \delta_{ij} + \bigl( 2 \g \frac{\kvec \cdot \Xvec}{|\kvec|^2} \bigr) \, k_i X_j + \bigl( - 2 \g \bigr) \, X_i X_j & \text{, \ $\Lscr_\mathrm{int} = \g \Ocal_4$} \\ 
    \delta_{ij} & \text{, \ $\Lscr_\mathrm{int} = \g \Ocal_5$} \\ 
    \end{cases} \\ 
    \mathbb{P}_{ij} & = \begin{cases}
    0 & \text{, \ $\Lscr_\mathrm{int} = \g \Ocal_1$} \\ 
    \bigl( 4 \g \Xvec \cdot \dot{\Xvec} \bigr) \, \delta_{ij} & \text{, \ $\Lscr_\mathrm{int} = \g \Ocal_2$} \\ 
    \bigl( -2 \g \bigr) \, \dot{X}_i X_j + \bigl( - 2 \g \bigr) \, X_i \dot{X}_j + \bigl( 2 \g \frac{\kvec \cdot \Xvec}{|\kvec|^2} \bigr) k_i \dot{X}_j + \bigl( 2 \g \frac{\kvec \cdot \dot{\Xvec}}{|\kvec|^2} \bigr) k_i X_j & \text{, \ $\Lscr_\mathrm{int} = \g \Ocal_3$} \\ 
    \bigl( 4 \g \Xvec \cdot \dot{\Xvec} \bigr) \, \delta_{ij} + \bigl( - 2 \g \bigr) \, \dot{X}_i X_j + \bigl( - 2 \g \bigr) \, X_i \dot{X}_j + \bigl( 2 \g \frac{\kvec \cdot \dot{\Xvec}}{|\kvec|^2} \bigr) \, k_i X_j + \bigl( 2 \g \frac{\kvec \cdot \Xvec}{|\kvec|^2} \bigr) \, k_i \dot{X}_j & \text{, \ $\Lscr_\mathrm{int} = \g \Ocal_4$} \\ 
    \bigl( - 2 i \g \kvec \cdot \dot{\Xvec} \bigr) \, \delta_{ij} & \text{, \ $\Lscr_\mathrm{int} = \g \Ocal_5$} \\ 
    \end{cases} \\ 
    \mathbb{Q}_{ij} & = \begin{cases}
    |\kvec|^2 \, \delta_{ij} + \bigl( 4 i \g \Xvec \cdot \dot{\Xvec} \bigr) \, \epsilon_{ijk} k_k & \text{, \ $\Lscr_\mathrm{int} = \g \Ocal_1$} \\ 
    |\kvec|^2 \, \delta_{ij} + \bigl( 2 \g |\kvec|^2 \, |\Xvec|^2 \bigr) \, \delta_{ij} & \text{, \ $\Lscr_\mathrm{int} = \g \Ocal_2$} \\ 
    |\kvec|^2 \, \delta_{ij} + \bigl( - 2 \g (\kvec \cdot \Xvec)^2 \bigr) \, \delta_{ij} + \bigl( - 2 \g |\kvec|^2 \bigr) \, X_i X_j + \bigl( 2 \g \kvec \cdot \Xvec \bigr) \, k_i X_j & \text{, \ $\Lscr_\mathrm{int} = \g \Ocal_3$} \\ 
    |\kvec|^2 \, \delta_{ij} + \bigl( - 2 \g (\kvec \cdot \Xvec)^2 + 2 \g |\kvec|^2 \, |\Xvec|^2 \bigr) \, \delta_{ij} + \bigl( - 2 \g |\kvec|^2 \bigr) \, X_i X_j + \bigl( 2 \g \kvec \cdot \Xvec \bigr) \, k_i X_j & \text{, \ $\Lscr_\mathrm{int} = \g \Ocal_4$} \\ 
    |\kvec|^2 \, \delta_{ij} + \bigl( - i \g \kvec \cdot \ddot{\Xvec} \bigr) \delta_{ij} & \text{, \ $\Lscr_\mathrm{int} = \g \Ocal_5$} \\ 
    \end{cases} 
    \;.
}
Here $k_i = (\kvec)_i$ and $X_i = [\Xvec(t)]_i$ and $A_i = [\Avec_\kvec(t)]_i$ with $\Avec(t,\xvec) = \int \! \mathrm{d}^3 \kvec \, \Avec_\kvec(t) \, e^{i \kvec \cdot \xvec}/ (2\pi)^3$.  
To derive these expressions we have made two simplifying assumptions.  
First, we work to leading order in powers of the coupling $\g$.  
If the dimensionless combination $\g |\Xvec|^2$ were to become $O(1)$, our EFT expansion would be invalid, and we are safe to assume $\g |\Xvec|^2 \ll 1$, which lets us work to leading order in $\g$.  
Second, we neglect gradients of the dark photon field.  
Whereas for a vector soliton, the dark photon field is inhomogeneous on a scale $\sim R$, the modes that exhibit parametric resonance are inhomogeneous on a much shorter length scale $\lambda = 2\pi/k \sim m^{-1} \ll R$.  
To study electromagnetic radiation in these modes and calculate their Floquet exponent, it is a good approximation to neglect spatial gradients of $\Xvec$~\cite{Hertzberg:2018zte}. 

\subsection{Applying Floquet theory}
\label{sub:floquet}

To identify the growing solutions of \eref{eq:OPQ_equation}, we adapt known techniques from Floquet theory. 
Floquet theory is well established, however, our system is somewhat non-trivial compared to the usual textbook examples because of the coupling of different components as well as constraints that must be respected.
Here, we follow sec.\,3.2.1-3.2.3 in Ref.~\cite{Amin:2014eta}, where a general framework to compute Floquet solutions was presented and is most easily adapted to our needs.  
\\ \\
\noindent{\bf Reduced system}: Before applying Floquet theory to analyse the solutions of \eref{eq:OPQ_equation},  we impose the Coulomb constraint $\kvec \cdot \Avec=0$ and eliminate $A_3$. Explicitly, $A_3=-k_3^{-1}(k_2A_2+k_1A_1)$ for $k_3\ne 0$. With this substitution, \eref{eq:OPQ_equation} becomes 
\begin{align}
\label{eq:tOPQ_equation}
\tilde{\mathbb{O}}_{ij}\ddot{A}_j+\tilde{\mathbb{P}}_{ij}\dot{A}_j+\tilde{\mathbb{Q}}_{ij}A_j=0,\qquad \textrm{where}\quad \tilde{\mathbb{O}}_{ij}\equiv \mathbb{O}_{ij}-\mathbb{O}_{i3}k_j/k_3\,\quad (\textrm{similarly for $\tilde{\mathbb{P}},\tilde{\mathbb{Q}}$ and  $i,j=1,2$}).
\end{align}
\Eref{eq:tOPQ_equation} is a system of two, second order differential equations which can be written as four first order equations:
\begin{equation}\label{eq:BigFloq}
    \dot{\bm{q}}(t) = \tilde{\mathbb{U}}(t) \, \bm{q}(t)
    \qquad \text{with} \qquad 
    \bm{q}(t)
    =
    \left( {\begin{array}{c}
    \Avec(t)\\
    \dot{\Avec}(t)
    \end{array} } \right)
    \quad\textrm{and}\quad 
    \tilde{\mathbb{U}}(t)
    =
    \left( {\begin{array}{c|c}
    0 & \mathds{1}\\\hline\\[-6pt]
    -\tilde{\mathbb{O}}^{-1}\tilde{\mathbb{Q}}& -\tilde{\mathbb{O}}^{-1}\tilde{\mathbb{P}} 
    \end{array} } \right) 
    \;.
\end{equation}
If $\tilde{\mathbb{U}}(t+T)=\tilde{\mathbb{U}}(t)$ is periodic with period $T$, then Floquet's theorem guarantees a general solution of the form $\bm{q}(t)=\sum_{s=1}^4 c_s \, {\bm{P}}_s(t) \, e^{\mu_st}$ where $\bm{P}_s(t+T)=\bm{P}_s(t)$, and $\mu_s$ are called the Floquet exponents.  
If $\Re[\mu_s]>0$ for any $s$, then the equation of motion admits exponentially growing solutions. 
\\ \\
\noindent{\bf Floquet Exponents}: 
The Floquet exponents may be calculated by solving the matrix equation $\dot{\mathbb{F}}(t)=\tilde{\mathbb{U}}(t) \, \mathbb{F}(t)$ with the initial condition $\mathbb{F}(0)=\mathds{1}$ (numerically if necessary). 
The matrix solution $\mathbb{F}(t)$ with this initial condition is often referred to as the fundamental solution. 
The fundamental solution evaluated at $t=T$ is called the Monodromy matrix $\mathbb{F}(T)$.  
Let $f_s=|f_s|e^{i\theta_s}$, with $s=1$ to $4$, be the (complex) eigenvalues of the Monodromy matrix $\mathbb{F}(T)$. 
Then, the Floquet exponents are given by $\mu_s=T^{-1}\left[\ln |f_s|+i\theta_s\right]$. Since ${\rm det}(\mathbb{F})=1$, it follows that $\sum_{s=1}^4\mu_s=0$.  
\\ \\
\noindent{\bf Fastest growing solutions}: 
Eigenvectors $\bm{\epsilon}_s$ of the Monodromy matrix provide the functions $\bm{P}_s(t)=\mathbb{F}(t)\bm{\epsilon}_se^{-\mu_st}$.  
Since $\bm{P}_s(t+T)=\bm{P}_s(t)$ is periodic, if one solves the equation numerically, a solution is only needed for one period (as is the case for calculating Floquet exponents). 
If we order the eigenvalues by the largest real part, then $\bm{q}_1(t)= c_1\bm{P}_1(t)e^{\mu_1 t}$ provides the fastest growing solution. 

So far we have suppressed the dependence of our quantities of interest on $\kvec$ to reduce clutter in the equations. 
Let us re-instate this dependence to discuss the fastest growing solutions more explicitly. 
For each Fourier mode, indexed by a wavevector $\kvec$, there are four Floquet exponents, and four eigenvectors corresponding to particular polarizations of the outgoing electromagnetic field.  
We label the Floquet exponents by $\mu_{\kvec,s}$ with arbitrary 3-vector $\kvec$ and with $s = 1-4$ (similarly for the eigenvectors $\bm{\epsilon}_{\kvec,s}$).
If the equation of motion admits exponentially growing solutions, the dynamics will be dominated by the solution that grows most quickly.  
Therefore it is useful to identify 
\ba{
    \mu_{\kvec,\mathrm{max}} = \max\limits_{s} \, \Re[\mu_{\kvec,s}]
    \qquad \text{and} \qquad 
    \mu_\mathrm{max} = \max\limits_{\kvec, \, s} \, \Re[\mu_{\kvec,s}]
    \;.
}
The quantity $\mu_{\kvec,\mathrm{max}}$ gives the largest real part of the four Floquet exponents for a given wavevector $\kvec$, and the quantity $\mu_\mathrm{max}$ gives the largest Floquet exponent among all possible wavevectors.  
In a given system, $\mu_\mathrm{max}$ parametrizes the growth rate of electromagnetic radiation, while $\mu_{\kvec,\mathrm{max}}$ parametrizes the radiation emitted in a particular direction and with a particular wavelength $\lambda = 2\pi / |\kvec|$.  
For a given $\kvec$, the polarization of the radiation is determined by inspecting $\bm{\epsilon}_{\kvec,\rm{max}}$, which denotes the eigenvector corresponding to the Floquet exponent with the largest real part for fixed $\bm{k}$. 
\\ \\
\noindent{\textbf{Analytical Approximations:}}
Since $\tilde{\mathbb{O}}^{-1}\tilde{\mathbb{Q}}$ and $\tilde{\mathbb{O}}^{-1}\tilde{\mathbb{P}}$ are periodic functions with period $T=2\pi/\omega_0$, they can be expanded as a Fourier series $\tilde{\mathbb{O}}^{-1}\tilde{\mathbb{Q}}=\sum_{l}[\tilde{\mathbb{O}}^{-1}\tilde{\mathbb{Q}}]_l e^{il\omega_0t}$ and $\tilde{\mathbb{O}}^{-1}\tilde{\mathbb{P}}=\sum_{l}[\tilde{\mathbb{O}}^{-1}\tilde{\mathbb{P}}]_l e^{il\omega_0t}$, where $l$ is an integer. 
In the small source amplitude regime, we expect a solution of the form $\tilde{\Avec}(t)=\tilde{\Avec}_+(t)e^{i\omega_0 t}+\tilde{\Avec}_-(t)e^{-i\omega_0 t}$, with a slowly varying $\tilde{\Avec}_{\pm}(t)$. 
Since interaction operators $\Ocal_1$ through $\Ocal_4$ are quadratic in the photon and dark photon fields, at leading order, this ansatz corresponds to the process $\Xvec+\Xvec\rightarrow \Avec+\Avec$. 
Here, the $\Xvec$ particles are at rest with initial energy $\omega_0$. The emitted photons  have the same energy $k+O(g^2)=\omega_0$.  
Plugging this ansatz in the reduced system of equations, and collecting terms $\propto e^{\pm i\omega_0 t}$, we arrive at 
\begin{equation}\label{eq:harmonictecnique}
    \dot{\tilde{\bm{y}}}(t)=\tilde{\mathbb{M}}\tilde{\bm{y}}(t)
    \quad \text{with} \quad 
\tilde{\bm{y}}(t)
=
\left( {\begin{array}{c}
\tilde{\Avec}_+(t)\\
\tilde{\Avec}_-(t)
 \end{array} } \right)\,,\qquad  
 \end{equation}
 where
  \begin{equation}
 \tilde{\mathbb{M}}=\left( {\begin{array}{c|c}
    
    -\frac{i(\omega_0^2-|{\kvec}|^2)}{2\omega_0}\mathds{1} - \frac{(\omega_0^2+|\kvec|^2)}{4\omega_0^2}[\tilde{\mathbb{O}}^{-1}\mathbb{\tilde{P}}]_0+\frac{i}{2\omega_0}\left[\overline{\mathbb{O}^{-1}\mathbb{Q}}   \right]_0
    &  \frac{(\omega_0^2+|{\kvec}|^2)}{4\omega_0^2}[\tilde{\mathbb{O}}^{-1}\tilde{\mathbb{P}}]_2+\frac{i}{2\omega_0}\left[  \tilde{\mathbb{O}}^{-1}\tilde{\mathbb{Q}}  \right]_2
    \\\hline\\[-6pt]
    
    \frac{(\omega_0^2+|{\kvec}|^2)}{4\omega_0^2}[\tilde{\mathbb{O}}^{-1}\tilde{\mathbb{P}}]_{-2}-\frac{i}{2\omega_0}\left[ \tilde{\mathbb{O}}^{-1}\tilde{\mathbb{Q}}  \right]_{-2}&
   
    \frac{i(\omega_0^2-|{\kvec}|^2)}{2\omega_0}\mathds{1} -  \frac{(\omega_0^2+|{\kvec}|^2)}{4\omega_0^2}[\tilde{\mathbb{O}}^{-1}\tilde{\mathbb{P}}]_0-\frac{i}{2\omega_0}\left[ \overline{\mathbb{O}^{-1}\mathbb{Q}}   \right]_0
   
    \end{array} } \right) 
    \;.
\end{equation}
Here, we have only kept terms up to $O(g^2)$ and use $\tilde{\mathbb{O}}^{-1}\tilde{\mathbb{P}}=O(g^2)$, $[\overline{\mathbb{O}^{-1}\mathbb{Q}}]_0=[\tilde{\mathbb{O}}^{-1}\tilde{\mathbb{Q}}]_0 - |\kvec|^2\mathds{1} = O(g^2)$. 
Note that these considerations mean that all entries in the above matrix are $O(g^2)$, and so are the eigenvalues.  
The four eigenvalues of $\tilde{\mathbb{M}}$ are the Floquet exponents $\mu_s$ for $s = 1-4$. 

\subsection{Linearly polarized dark photon field}
\label{sub:linear_pol}

We consider a homogeneous and linearly-polarized dark photon field, which is written as 
\ba{\label{eq:linearsoliton}
    \Xvec(t,\xvec) = \Xbar \, \cos(m t) \, \hat{\bm z} 
    \;,
}
where $\Xvec$ has a constant orientation and varying magnitude. We have set the temporal oscillation frequency $\omega_0=m$ which is an excellent approximation in the non-relativistic limit.
For each of the operators, $\mathcal{O}_1$ through $\mathcal{O}_5$, we perform the Floquet analysis described above, working to leading order in powers of the coupling $\g$.  
To illustrate the details of these analytic calculations, we work through the derivation for operator $\Ocal_3$ in \aref{apphomo:linear}; the calculations for other operators are similar.  
For each operator, the maximum Floquet exponent (real part) is found to be 
\ba{\label{eq:muk_linear}
    \mu_\mathrm{max}  
    & = \begin{cases}
    \frac{1}{2} \g \Xbar^2 m & , \quad \text{for $\Lscr_\mathrm{int} = g^2\Ocal_1$} \\ 
     \frac{1}{2} \g \Xbar^2 m & , \quad \text{for $\Lscr_\mathrm{int} = g^2\Ocal_2$} \\ 
    \frac{1}{2} \g \Xbar^2 m & , \quad \text{for $\Lscr_\mathrm{int} = g^2\Ocal_3$} \\ 
    \frac{1}{2} \g \Xbar^2 m & , \quad \text{for $\Lscr_\mathrm{int} = g^2\Ocal_4$} \\ 
    O(g^4) & , \quad \text{for $\Lscr_\mathrm{int} = g^2\Ocal_5$} \\ 
    \end{cases} 
    \;,
}
where the maximization is performed over all possible wavevectors $\kvec$ and all possible polarizations of the outgoing radiation.  
The results are equivalent for operators $\Ocal_1$ through $\Ocal_4$, and we discuss these results further below.  
For $\Ocal_5$, the real part of the Floquet exponent is parametrically higher order in the coupling.  
This is because the additional time derivative in $\Ocal_5$, see \eref{eq:operators}, brings a factor of $i$ which renders the leading-order Floquet exponent imaginary.  
 
\begin{figure}[t]
\centering
  \includegraphics[scale=0.5]{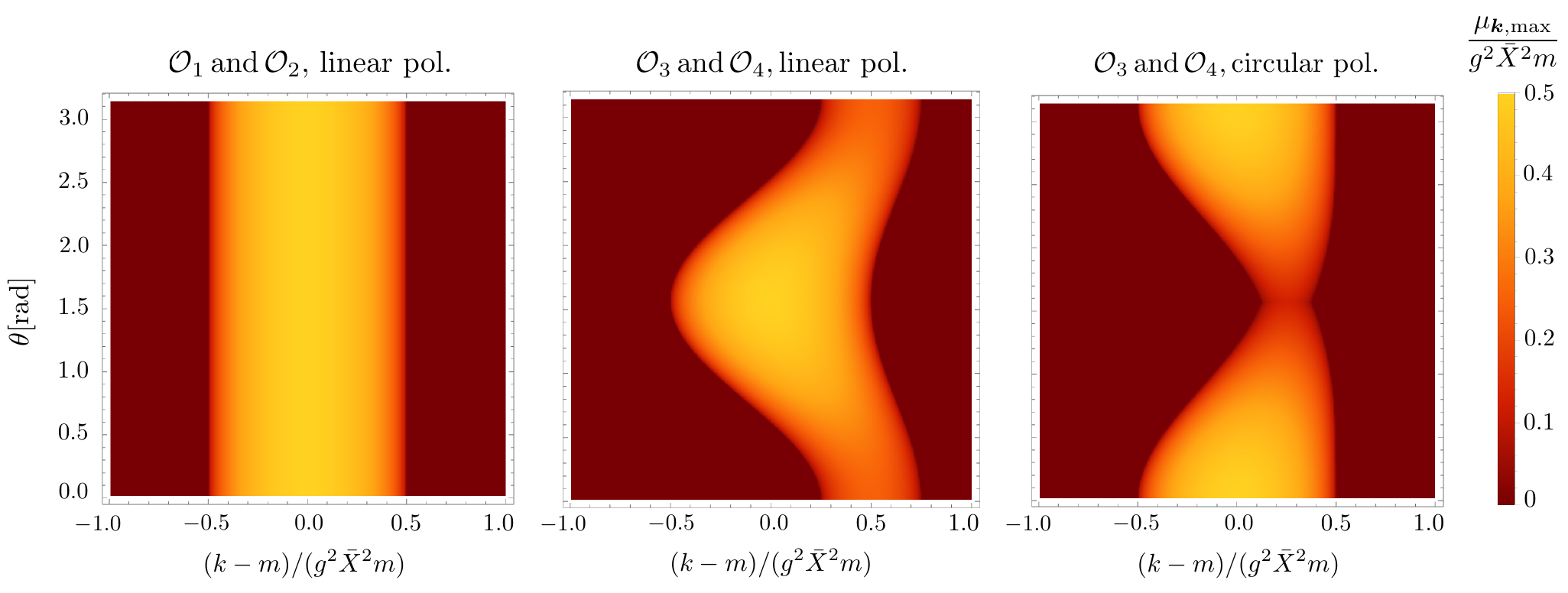}
  \includegraphics[scale=0.45]{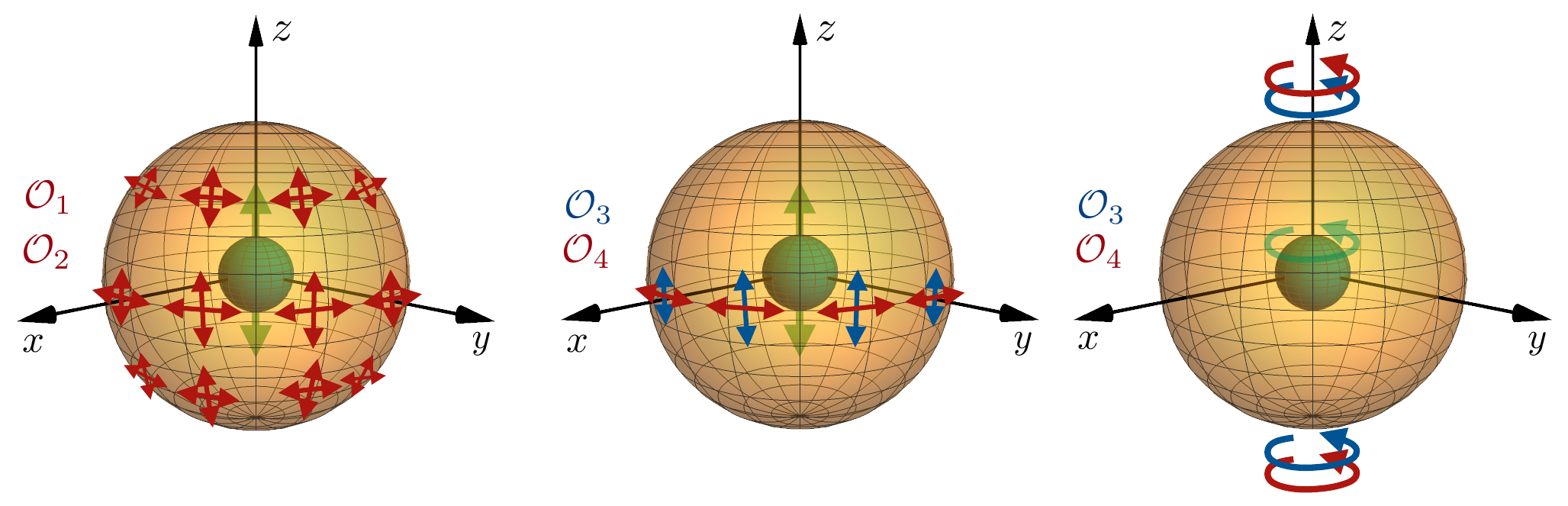}
\caption{
\label{Num:muktheta}
Electromagnetic radiation arising from a homogeneous dark photon field coupled to electromagnetism though several dimension-6 operators via the phenomenon of parametric resonance.  
\textit{Top:}  The maximal Floquet exponent $\mu_{\kvec,\mathrm{max}}$ is shown as a function of the wavenumber $k$ of the electromagnetic radiation and the polar angle $\theta$ such that $\cos \theta = \kvec \cdot \hat{\bm z} / k$.  The dominant Floquet band is centered at $k \approx m$ and has width $O(g^2 \Xbar^2 m)$, where $m$ is the dark photon mass, $\Xbar$ is the field amplitude, and $g$ is the coupling to electromagnetism with $\Lscr_\mathrm{int} = g^2 \Ocal_i$.  The three panels correspond to different operators $\Ocal_i$ and different polarizations for the dark photon field. 
\textit{Bottom:}  These graphics illustrate the orientation of the radiation's polarization.  The green arrows denote the polarization of the dark photon field (e.g., vector soliton), while the red and blue arrows denote the polarization of the emitted radiation (for different operators).  For operators $\Ocal_1$ and $\Ocal_2$ (bottom-left) the radiation is emitted isotropically, and has no preferred polarization direction.  For operators $\Ocal_3$ and $\Ocal_4$ with a linearly-polarized dark photon field (bottom-middle) the radiation is predominatly emitted in the directions normal to $\hat{\bm z}$, whereas for a circularly-polarized dark photon field (bottom-right) the emission is predominantly aligned with $\pm \hat{\bm z}$. }
\end{figure}

 First we discuss operators $\Ocal_1$ and $\Ocal_2$. For both of these operators, the dark photon field enters via $\Xvec \cdot \Xvec$, so its indices are not `entangled' with the electromagnetic field.  
Consequently both $\Ocal_1$ and $\Ocal_2$ have the same behavior in regard to the direction and polarization of the radiation.  
We find that $\mu_{\kvec,\mathrm{max}}$ is independent of the wavevector's orientation, and the electromagnetic radiation is emitted isotropically.  
Since the operators only depend on $|\Xvec|$, the radiation doesn't `know' about the dark photon field's orientation, and we obtain the same radiation pattern as if the condensate had been a scalar field~\cite{Hertzberg:2018zte}.  
Our numerical results for operator $\Ocal_2$ are illustrated in the top-left panel of \fref{Num:muktheta}, and the chart for $\mathcal{O}_1$ is indistinguishable. The dominant Floquet band is centered at $k = m$.  
The isotropic emission is reflected in the `vertical' nature of the Floquet band, which is independent of the angle $\theta$ between $\kvec$ and $\Xvec$. For both operators, the emitted radiation has no preferred polarization direction as shown in the left bottom panel in \fref{Num:muktheta}. 

Next we discuss operators $\Ocal_3$ and $\Ocal_4$.  
Here the indices for the dark photon field contract with the indices for the electric and magnetic fields, and this leads to a richer structure in the Floquet chart.  
The top-middle panel of \fref{Num:muktheta} shows the Floquet charts for $\Ocal_3$ and $\Ocal_4$ which are identical.  
The maximal Floquet exponent $\g \Xbar^2 m / 2$ is obtained for $\theta = \pi/2$, corresponding to emission that is normal to the dark photon field's orientation, $\kvec \perp \hat{\bm z}$. 
Whereas for $\theta = 0$ or $\pi$, corresponding to $\kvec = k_z \, \hat{\bm z}$, the Floquet exponent is smaller by a factor of $2$.  
More generally, our analytical analysis yields an expression~(\ref{eq:muA303linear}) for the maximal Floquet exponent (maximizing over orientations of the electromagnetic field's polarization) with an arbitrary angle $\theta$ between $\kvec$ and $\hat{\bm z}$, which is given by 
\ba{
    \mu_{\kvec,\mathrm{max}}(\theta) = \frac{1}{2} \g \Xbar^2 m \left( 1 - \frac{1}{2} \cos^2 \theta \right) 
    \;.
}
The radiation's polarization is found to be different for the two operators.  
For operator $\Ocal_3$ the outgoing radiation at $\theta = \pi/2$ is polarized in the direction of the dark photon field $\Xvec \propto \hat{\bm z}$, and for operator $\Ocal_4$ it is normal to the dark photon field in the azimuthal direction $\hat{\bm \phi}$, as indicated in the bottom-middle panel of \fref{Num:muktheta}.  

\subsection{Circularly polarized dark photon field}
\label{sub:circular_pol}

We consider a homogeneous and circularly-polarized dark photon field, which is written as 
\ba{\label{eq:HCcirularly}
    \Xvec(t,\xvec) = \frac{\Xbar}{\sqrt{2}} \, \bigl( \cos(m t) \, \hat{\bm x} + \sin(m t) \, \hat{\bm y} \bigr)
    \;,
}
where $\Xvec$ has a constant magnitude and varying orientation.  
By performing the Floquet analysis described above, we calculate the Floquet exponents $\mu_{\kvec,s}$.  
We provide some details of this derivation for $\Ocal_3$ in \aref{Sec:O3circular}.  
Maximizing the real part over all possible directions and polarizations of the outgoing radiation yields 
\ba{\label{eq:muk_circular}
    \mu_\mathrm{max} 
    & = \begin{cases}
    0 & , \quad \text{for $\Lscr_\mathrm{int} = g^2\Ocal_1$} \\ 
    0 & , \quad \text{for $\Lscr_\mathrm{int} = g^2\Ocal_2$} \\ 
    \frac{1}{2} \g \Xbar^2 m & , \quad \text{for $\Lscr_\mathrm{int} = g^2\Ocal_3$} \\ 
    \frac{1}{2} \g \Xbar^2 m & , \quad \text{for $\Lscr_\mathrm{int} = g^2\Ocal_4$} \\ 
    O(g^4) & , \quad \text{for $\Lscr_\mathrm{int} = g^2\Ocal_5$} 
    \end{cases} 
    \;.
}
Operators $\Ocal_1$ and $\Ocal_2$ do not lead to parametric resonance for a circularly polarized dark photon field, hence $\mu_\mathrm{max} = 0$.  
For these operators, the dark photon field enters through $|\Xvec|$, which remains constant in the circularly-polarized configuration~(\ref{eq:HCcirularly}).  
For operator $\Ocal_5$, the Floquet exponent is imaginary at $O(\g)$; see \sref{sub:linear_pol}.  

Next we discuss operators $\Ocal_3$ and $\Ocal_4$.  
The analytic calculations are facilitated by moving to a circular polarization basis for the outgoing radiation.  
The top-right panel of \fref{Num:muktheta} shows the Floquet chart for operator $\Ocal_3$, and the chart for $\Ocal_4$ is indistinguishable.  
The Floquet exponent is maximized for $\theta = 0$ and $\pi$, corresponding to radiation in the direction normal to the plane of the dark photon field, $\kvec = k_z \, \hat{\bm z}$, as shown in the right bottom panel of \fref{Num:muktheta}.  
The radiation carries circular polarization with the same handedness as the dark photon field.  
This means that the radiation emitted from $\theta = 0$ and $\theta = \pi$ have opposite helicity.  

\section{Radiation from polarized vector solitons}
\label{sec:vector_soliton}

In this section we adapt the results of our Floquet analysis to study electromagnetic radiation from polarized vector solitons.  

\subsection{Condition for parametric resonance}
\label{sub:condition}

Our Floquet analysis is performed assuming a homogeneous dark photon field $\Xvec(t,\xvec) = \Xvec(t)$.  
Of course, a vector soliton is not a homogeneous field configuration; rather, the field's amplitude drops smoothly to zero beyond a distance $r \approx R$ away from the soliton's center.  
Nevertheless, earlier work~\cite{Hertzberg:2018zte} has established that for scalar solitons the maximal Floquet exponent is insensitive to the soliton's finite size provided that the soliton is sufficiently large.  
Here we show that these arguments carry over to vector solitons as well.  
Specifically, we claim that the maximum Floquet exponent $\mu_\mathrm{max}^\mathrm{(sol.)}$ of the electromagnetic radiation emitted by a polarized vector soliton can be approximated by 
\ba{\label{eq:muk_approx}
    \mu_\mathrm{max}^\mathrm{(sol.)} \approx \begin{cases} 
    \mu_\mathrm{max}^\mathrm{(hom.)} - R^{-1} & , \quad \mu_\mathrm{max}^\mathrm{(hom.)} R \gtrsim 1 \\ 
    0 & , \quad \mu_\mathrm{max}^\mathrm{(hom.)} R \lesssim 1 
    \end{cases} 
}
where $\tau_\text{\sc lc} \approx 2R$ is the light-crossing time of a soliton with radius $R$, and we have dropped a factor of order unity
(see App.\,\ref{app:spherical} and Fig.\,\ref{ProofRes}).  
In this relation $\mu_\mathrm{max}^\mathrm{(hom.)}$ is the maximal Floquet exponent in a homogeneous system with $\Xvec(t,\xvec) = \Xvec(t,0)$ equal to the dark photon field at the soliton's center.  
We have already presented results for $\mu_\mathrm{max}^\mathrm{(hom.)}$ assuming that the homogeneous dark photon field is either linearly or circularly polarized; see \erefs{eq:muk_linear}{eq:muk_circular}. 
We motivate the approximation in \eref{eq:muk_approx} by directly calculating the Floquet exponent using a spherical soliton profile for operator $\Ocal_1$; we present these results in \aref{app:spherical}. 

The condition $\mu_\mathrm{max}^\mathrm{(hom.)} R > 1$ must be satisfied in order for parametric resonance to occur.
Since $\mu_\mathrm{max}^\mathrm{(hom.)}$ is the instability growth rate and $R$ is the soliton's light-crossing time, this condition expresses the fact that radiation is being generated more quickly than it is leaving the system, and parametric resonance results from the associated Bose enhancement. 
This condition imposes a lower limit on the strength of the coupling.  
Using the expression for $\Xbar$ from \eref{eq:Psia} and the expression for $R$ from \eref{eq:EMR}, the parametric resonance condition is expressed as 
\ba{\label{eq:PR_condition}
    \g \Xbar^2 > \Bigl( \frac{\mu}{m} \Bigr)^{1/2}
    \qquad \text{and} \qquad 
    g \mpl > \Bigl( \frac{\mu}{m} \Bigr)^{-3/4}
}
where $\mu$ is the chemical potential for the soliton solution.  
In addition, the coupling must remain small to justify truncating the EFT at dimension-6 operators; see \eref{eq:EFT_condit}.  
Taken together, the conditions for valid EFT and successful parametric resonance imply $(\mu/m)^{1/2} < \g \Xbar^2 \ll 1$ and $(\mu/m)^{-3/4} < g \mpl \ll (\mu/m)^{-1}$.  
Both conditions can be satisfied provided that $\mu/m \ll 1$.  
Recall that $\mu/m < 1$ is required for validity of the non-relativistic expansion, and $\mu/m \ll 1$ is typical for soliton solutions. 

It is instructive to compare the above resonance condition with the case when we have a scalar soliton. The resonance phenomenon for scalar (or pseudoscalar) solitons is different in several aspects from that for vector solitons. 
The spatially-localized configuration with spherically-symmetric density profiles for such solitons takes the form $\phi(t, r) = (\sqrt{2/m})\psi(r) \text{cos}(\omega t)$, where $\psi(r)$ is well-approximated by the empirical fitting formula given in eq.\,(\ref{eq:Psia}), under the replacement $\bar{X} \rightarrow \bar{\phi}$. The dimension-5 operators which enter in play are $\bar{\mathcal{O}}_1=-(1/4)F_{\mu\nu}\tilde{F}^{\mu\nu}\phi$ and $\bar{\mathcal{O}}_2=-(1/4)F_{\mu\nu}F^{\mu\nu}\phi$, where  $\Lscr_{\text{int}}= g \bar{\mathcal{O}}_1$ and $\Lscr_{\text{int}}= g \bar{\mathcal{O}}_2$, respectively.  The resonance condition takes the same form as eq.\,(\ref{eq:muk_approx}), but now $\mu_{\text{max}}^{\text{(hom.)}}$ is the maximal Floquet exponent in a homogeneous system with $\phi(t,\xvec)=\phi(t,0)$ equal to the scalar field at the soliton's center. Since only one scalar connects to two photons, the maximal Floquet exponent for both operators is proportional to only one power of the product between the constant field amplitude and the coupling constant. Moreover, the dominant Floquet band is centered at $k=m/2$.  For both operators, we have $\mu_{\text{max}}^{\text{(hom.)}} = g \bar{\phi} m/4$. As a result, the parametric resonance condition is expressed as (again ignoring factors of order unity):
$
    g \bar{\phi} > ( {\mu}/{m})^{1/2}
    \qquad \text{and} \qquad 
    g \mpl > ({\mu}/{m})^{-1/2}$
where $\mu$ is the chemical potential of the scalar soliton solution. For the same value of $\mu/m$ for scalar and vector solitons, the resonance phenomenon requires a larger $g$ in vector solitons compared to the scalar case. Conversely for the same $g$, the resonance condition can be satisfied for larger values of $\mu/m$ for scalars compared to vectors (ie. for fixed $m$, smaller radii solitons).

\subsection{Vector soliton decay}

Electromagnetic radiation via parametric resonance extracts energy from the vector soliton.  
If this emission continues for a sufficiently long time, it would eventually cause the vector soliton to decay.  
For each of the dimension-6 operators, and for both the linearly- and circularly-polarized soliton configurations, we estimate the vector soliton's lifetime as $\tau = 1 / \mu_\mathrm{max}^\mathrm{(sol.)} \approx 1 / \mu_\mathrm{max}^\mathrm{(hom.)}$. 
Assuming that the condition for parametric resonance (\ref{eq:PR_condition}) is satisfied, and using the results in \erefs{eq:muk_linear}{eq:muk_circular}, the soliton lifetime is calculated as 
\ba{\label{eq:tau}
    \tau & = \begin{cases}
    2 / \g \Xbar^2 m & , \quad \text{for $\Lscr_\mathrm{int} = \g \Ocal_1$} \\ 
    2 / \g \Xbar^2 m & , \quad \text{for $\Lscr_\mathrm{int} = \g \Ocal_2$} \\ 
    2 / \g \Xbar^2 m & , \quad \text{for $\Lscr_\mathrm{int} = \g \Ocal_3$} \\ 
    2 / \g \Xbar^2 m & , \quad \text{for $\Lscr_\mathrm{int} = \g \Ocal_4$} \\ 
    O(g^{-4}) & , \quad \text{for $\Lscr_\mathrm{int} = \g \Ocal_5$} 
    \end{cases} 
    \qquad \text{and} \qquad 
    \tau = \begin{cases}
    \infty & , \quad \text{for $\Lscr_\mathrm{int} = \g \Ocal_1$} \\ 
    \infty & , \quad \text{for $\Lscr_\mathrm{int} = \g \Ocal_2$} \\ 
    2 / \g \Xbar^2 m & , \quad \text{for $\Lscr_\mathrm{int} = \g \Ocal_3$} \\ 
    2 / \g \Xbar^2 m & , \quad \text{for $\Lscr_\mathrm{int} = \g \Ocal_4$} \\ 
    O(g^{-4}) & , \quad \text{for $\Lscr_\mathrm{int} = \g \Ocal_5$} 
    \end{cases} 
}
for the linearly-polarized and circularly-polarized vector solitons, respectively.  
Recall that operators $\Ocal_1$ and $\Ocal_2$ do not lead to electromagnetic radiation from a circularly-polarized vector soliton, and we write $\tau = \infty$.  

With these estimates, we turn to the question of vector soliton stability and decay.  
Since $T \approx 2\pi/m$ is the oscillation period of the non-relativistic dark photon field, and for $\g \Xbar^2 \ll 1$, these formulas reveal that the vector soliton survives for many cycles of oscillation, $\sim \tau / T \approx 1/\g \Xbar^2 \gg 1$.  
However, for parameters that are typical of dark photon dark matter, $m^{-1} \sim (10^{-6} \, \mathrm{eV})^{-1} \sim 10^{-9} \, \mathrm{s}$, the lifetime $\tau$ is still very short compared to the age of the universe today $t_0 \sim 10^{17} \, \mathrm{s}$.  
The conclusion is that any solitons in the Universe today must fail to meet the parametric resonance condition (\ref{eq:PR_condition}), which shuts off the channel for their decay into electromagnetic radiation.  
The observation has been noted previously for axion dark matter with a dimension-5 coupling to electromagnetism~\cite{Hertzberg:2018zte, Hertzberg:2020dbk}.  

The condition for soliton stability is the converse of the condition for parametric resonance (\ref{eq:PR_condition}).  
A stable soliton must have a weak coupling to electromagnetism such that $g \mpl \lesssim (\mu/m)^{-3/4}$.  
Using \eref{eq:EMR} this condition is expressed as an upper limit on the mass of the soliton: 
\ba{\label{eq:Mc}
    M 
    \lesssim 
    M_c \equiv 
    \frac{10^2 \mpl^{4/3}}{g^{2/3} m} 
    \sim \bigl( 3 \times 10^{21} \ \mathrm{kg} \bigr) \Bigl( \frac{m}{10^{-6} \ \mathrm{eV}} \Bigr)^{-1} \Bigl( \frac{g}{10^{-10} \ \mathrm{GeV}^{-1}} \Bigr)^{-2/3} 
    \;.
}
For these fiducial parameters, we also have $\Xbar \sim (1 \times 10^{8} \ \mathrm{GeV}) \, g_{10}^{-4/3}$, $N \sim (2 \times 10^{63}) g_{10}^{-2/3} m_6^{-2}$, and $R \sim (100 \ \mathrm{km}) g_{10}^{2/3} m_6^{-1}$ where $m_6 \equiv m / 10^{-6} \ \mathrm{eV}$ and $g_{10} \equiv g / 10^{-10} \ \mathrm{GeV}^{-1}$.  
\Eref{eq:Mc} gives an upper limit on the mass of polarized vector solitons that we should expect to find in the Universe today.  

Finally let us address the cosmological history of vector soliton decay.  
At the time of soliton formation, there maybe be solitons with $M > M_c$.  
The parametric resonance is ineffective until the age of the Universe reaches $\tau$, and subsequently these solitons begin to decay.  
However, their decay is halted when $M$ decreases below $M_c$ and the channel for parametric resonance is blocked.  
Consequently, we expect that any solitons formed with $M > M_c$ should have $M \approx M_c$ today, just below the threshold for parametric resonance.  
A similar cosmological evolution has been discussed previously in the context of scalar solitons coupled to electromagnetism~\cite{Hertzberg:2020dbk}.  

\subsection{Astrophysical signatures from soliton mergers}

In light of the discussion from the preceding section, isolated vector solitons in the Universe today are not expected to produce electromagnetic radiation since the condition for parametric resonance is not met: $M < M_c$.  
However, it is reasonable to expect that an appreciable population of vector solitons with masses just below the threshold for parametric resonance may reside in the Milky Way halo.  
The merger of these sub-critical vector solitons may trigger a burst of electromagnetic radiation.  
This radiation can be understood to arise from a temporary `activation' of parametric resonance when the mass of the merged pair exceeds the threshold: $M_1 + M_2 > M_c$ although $M_1, M_2 < M_c$.  

The collision and merger of two solitons is a complicated non-linear process, and it is challenging to obtain accurate predictions with analytical methods.  
Nevertheless, 3-dimensional simulations have been performed using numerical lattice techniques; see Refs.~\cite{Schwabe:2016rze,Hertzberg:2020dbk,Amin:2020vja} for work on scalar solitons and Ref.~\cite{Amin:2022pzv,Jain:2022agt} for work on vector solitons.  
For both the scalar and vector soliton studies, the collision induces radiation that carries away an $O(1)$ fraction of the constituent particles and mass, leaving an approximately spherically-symmetric condensate.  
The simulations reported in Refs.~\cite{Schwabe:2016rze,Hertzberg:2020dbk} exhibit a mass for the merged system that is approximately $M_\mathrm{final} \approx 0.7 (M_1 + M_2)$ in terms of the progenitor masses (similar results were also seen for vector solitons in \cite{Amin:2022pzv}). 
This relation would allow $M_\mathrm{final} > M_c$ while $M_1, M_2 < M_c$, meaning that the merger could `trigger' parametric resonance.  
Although it is worth noting that these simulations do not allow for the coupling to electromagnetism that we study here, and any potential back reaction effects have been neglected.

For collisions of vector solitons, the polarization state (or spin density) of the transient and final dark photon configuration can impact the electromagnetic signatures from the merger. 
Our calculations have assumed maximally-polarized configurations with spin $|{\bm S}|$ = $0$ and $\hbar N$ in the linearly- and circularly-polarized configurations, respectively.  
During the merger process, it is likely that the system is better characterized as a fractionally polarized soliton~\cite{Jain:2021pnk} with $0<|{\bm S}|<\hbar N$ or as an excited, non-solitonic state. 
It is straightforward to extend our calculation to fractionally polarized solitons; however, we have not attempted to characterize excited states and their signatures. 
Simulations of collisions and mergers will help to reveal the realistic range of initial conditions for parametric resonance, leading to more robust predictions for the associated electromagnetic radiation.  

We are interested in the spectrum of electromagnetic radiation resulting from vector soliton collisions.  
For these estimates, we model the radiation as `triggered' parametric resonance. 
That is to say, once the merger has `completed' the radiation is emitted as a sudden burst that carries away an $O(1)$ fraction of the excess mass $\Delta M = M_\mathrm{final} - M_c$.  
These dynamics have been observed previously in simulated collisions of non-gravitationally-bounded scalar solitons with a coupling to electromagnetism~\cite{Amin:2022pzv}.  
However, our interest is in gravitationally-bounded soliton solutions for which the soliton's light-crossing time scale $R$ is many orders of magnitude larger than the time scale for parametric resonance $\tau$.  
For such solutions, it is possible that the radiation seeps out via less-intense bursts as the merged configuration settles into a spherically-symmetric condensate~\cite{Levkov:2020txo}.  
This discussion motivates further study of gravitationally-bounded vector soliton collisions.  
We expect that our approach, the `triggered' parametric resonance model, overestimates the strength of the signal, since it allows for the largest possible energy release and the smallest possible emission duration.  

To characterize the astrophysical signature associated with such a phenomenon, we need the signal duration $\tau$, the central wavelength $\lambda_0$, and the signal bandwidth $\Delta \lambda$. 
The soliton lifetime $\tau$ from \eref{eq:tau} sets the signal duration.  
For a fiducial set of parameters, we estimate 
\ba{
    \tau \sim \bigl( 20 \ \mu\mathrm{s} \bigr) \Bigl( \frac{g}{10^{-10} \ \mathrm{GeV}^{-1}} \Bigr)^{2/3} \Bigl( \frac{m}{10^{-6} \ \mathrm{eV}} \Bigr)^{-1} 
    \;,
}
where the critical mass condition $M = M_c$ has been used to eliminate $\mu$.  
The signal wavelength $\lambda_0 = 2\pi / k_0$ is controlled by the wavenumber $k_0 \approx m$ at the first instability band of the Floquet chart, and the signal bandwidth $\Delta \lambda = (2 \pi / k_0^2) \Delta k$ is controlled by the width of the Floquet band $\Delta k \approx \g \bar{X}^2 m$.  
For the fiducial parameters we estimate the corresponding frequencies to be 
\bsa{}{
    \nu_0 & \sim \bigl( 200 \ \mathrm{MHz} \bigr) \Bigl( \frac{m}{10^{-6} \ \mathrm{eV}} \Bigr)^{} \;, \\ 
    \Delta \nu & \sim \bigl( 40 \ \mathrm{kHz} \bigr) \Bigl( \frac{g}{10^{-10} \ \mathrm{GeV}^{-1}} \Bigr)^{-2/3} \Bigl( \frac{m}{10^{-6} \ \mathrm{eV}} \Bigr)^{} 
    \;.
}
These estimates imply that the radiation will be nearly monochromatic (a consequence of $\g \Xbar^2 \ll 1$).  
For the fiducial mass parameter $m = 10^{-6} \ \mathrm{eV}$, the emission is in the radio band of the electromagnetic spectrum.  
Since radio telescopes lose sensitivity below a frequency of $\sim (10-15) \ \mathrm{MHz}$, due to absorption and scattering in the ionosphere, only models with $m \gtrsim 5 \times 10^{-8} \ \mathrm{eV}$ could be probed with ground-based radio observations.\footnote{This issue also arises in axion search strategies such as that related to the axion-photon conversion during axion ultracompact minihalo-neutron star encounters (see, for example, Sec.\,VII in Ref.~\cite{Choi:2022btl}). One solution is to consider planned space-based facilities such as the Orbiting Low Frequency Antennas for Radio Astronomy Mission (OLFAR)~\cite{Bentum:2020}.}  

The strength of the signal is parametrized by a spectral flux density $S_B$.  
If the source is a distance $d$ away and it liberates an energy $O(M_c)$ in a time $\tau$, then we can estimate 
\ba{\label{eq:SB}
    S_B 
    & \sim \frac{M_c/\tau}{4\pi\Delta\nu d^2} 
    \sim \bigl( 3 \times 10^{18} \ \mathrm{Jy} \bigr) \Bigl( \frac{g}{10^{-10} \ \mathrm{GeV}^{-1}} \Bigr)^{-2/3} \Bigl( \frac{m}{10^{-6} \ \mathrm{eV}} \Bigr)^{-1} \Bigl( \frac{d}{1 \ \mathrm{Mpc}} \Bigr)^{-2} 
    \;,
}
where $1 \ \mathrm{Jy} = 10^{-26} \, \mathrm{W}/\mathrm{m}^2/\mathrm{Hz}$.  
For a cosmologically-distant source, the effect of cosmological redshift must also be included. Note that reducing the coupling $\g$ increases the strength of the signal, since the suppression from $(\Delta \nu)^{-1} \sim g^{2/3}$ is counterbalanced by the enhancements from $\tau^{-1} \sim g^{-2/3}$ and
$M_c \sim g^{-2/3}$, so that more energy per unit of frequency is emitted over a shorter time. 

Radio telescopes typically have sensitivities at the level of $\sim 100 \ \mu\mathrm{Jy}$ at $100 \ \mathrm{kHz}$ with $\sim 1 \ \mathrm{kHz}$ resolution bandwidth \cite{SKA1}.  
Our estimate in \eref{eq:SB} suggests that if a vector soliton collision triggers parametric resonance while the host galaxy is being observed, then the signal would easily be detectable.  
However, we must remember that our model generously overestimates the energy liberated and underestimates the duration of release.  
One should not interpret \eref{eq:SB} as a prediction for the spectral flux density, but rather an indication from dimensional analysis that the signal may be strong enough to detect.   

Radio telescopes, such as Green Bank Telescope (GBT), measure not just the intensity but also the polarization of incident radio waves \cite{2017isra.book.....T}. 
A measurement of the polarization would prove invaluable to discriminate among different possible soliton sources.  
Whereas a scalar soliton emits unpolarized radiation, a vector soliton, such as the ones we study here, may produce polarized radiation.  
In this way, a detection of polarized emission could be interpreted as evidence of vector soliton mergers.  
Moreover, the polarization strength and orientation depends on the nature of the coupling between the dark photon field and electromagnetism, providing an additional handle on the underlying particle physics model.  

However, a realistic analysis of the expected polarization signal is non-trivial. 
First, depending on the particular UV embedding, we expect that several dimension-6 operators would simultaneously source resonance. 
Each may lead to a different polarization pattern for the resultant radiation. 
Second, individual solitons or solitons produced from mergers may be fractionally polarized, as discussed already above \cite{Amin:2022pzv}. 
This too would complicate the resultant polarization pattern.  

\section{Summary and conclusion}
\label{sec:Conclusion}

In this work we have studied the electromagnetic radiation that arises via parametric resonance from a spatially-coherent dark photon field that interacts with the electromagnetic field via several dimension-6 operators.  
We study a homogeneous field and adapt these results to assess the radiation from polarized vector solitons formed from dark photon dark matter.  
The calculations presented in this article represent predictions for the electromagnetic signals arising from polarized vector solitons, and provide an avenue for probing soliton collisions and mergers. 

We identify five dimension-6 operators that couple a massive dark photon field to electromagnetism and lead to parametric resonance.  
These operators take the form $X_\alpha X_\beta F_{\mu\nu} F_{\rho\sigma}$ and $\partial_\alpha X_\beta F_{\mu\nu} F_{\rho\sigma}$.  
There also exist dimension-6 operators with one fewer factor of the photon field, which we do not consider here, since they do not lead to parametric resonance.  
We consider systems with either a linearly-polarized or a circularly-polarized dark photon field.  
In the first scenario, the orientation of the dark photon field $\Xvec(t,\xvec)$ remains fixed as its magnitude oscillates, whereas in the second scenario, the orientation oscillates while the magnitude remains fixed.  
For each of the five operators and both of the polarization configurations, we perform a Floquet analysis using both analytical and numerical methods.  
The electromagnetic field is exponentially amplified via parametric resonance with $|\Avec(t)| \propto e^{\mu_\mathrm{max} t}$.  
We calculate the maximal Floquet exponents $\mu_\mathrm{max}$ assuming a dark photon field with either linear or circular polarization, and these results are summarized below.  
\begin{itemize}
\item  For a linearly-polarized 
    vector soliton, operators of the form $X_\alpha X_\beta F_{\mu\nu} F_{\rho\sigma}$ lead to a maximum resonance growth rate (Floquet exponent) that is parametrically $\sim \g \Xbar^2 m$, where $\g$ is the dimension-6 operator coefficient, $\Xbar$ is the amplitude of the dark photon field, and $m$ is the dark photon mass.  This result agrees parametrically with earlier work on axion dark matter (spin-0 particles)~\cite{Hertzberg:2018zte} where the interaction is $\phi F_{\mu\nu} \tilde{F}^{\mu\nu}$,  where $\phi$ is the scalar field amplitude.  We find that the resonance band is centered at a wavenumber of $|\kvec| = m$, which sets the frequency of the resultant electromagnetic radiation.  By contrast, the axion case gives $|\kvec| = m/2$. For operators $X_\alpha X^\alpha F_{\mu\nu} F^{\mu\nu}$ and $X_\alpha X^\alpha F_{\mu\nu} \tilde{F}^{\mu\nu}$, the emitted radiation does not show any preferred polarization orientation.  For operators $X^\mu X_\nu F_{\mu\rho} F^{\nu\rho}$ and $X^\mu X_\nu F_{\mu\rho} \tilde{F}^{\nu\rho}$, we find that the radiation is primarily linearly polarized  along an axis that differs for each interaction operator. In this case, outgoing radiation is peaked in the equatorial plane perpendicular to the direction of oscillation of the dark photon field.
    \item  For a circularly-polarized  vector soliton, operators $X_\alpha X^\alpha F_{\mu\nu} F^{\mu\nu}$ and $X_\alpha X^\alpha F_{\mu\nu} \tilde{F}^{\mu\nu}$ do not lead to electromagnetic radiation via parametric resonance.  This is because $X_\alpha X^\alpha = -(X_0)^2 + \Xvec \cdot \Xvec$ is static for a circularly-polarized dark photon field.  Other operators lead to parametric resonance with a growth rate that is parametrically $\sim \g \Xbar^2 m$, similar to the linearly-polarized scenario.  We find that the outgoing radiation is primarily circularly polarized with the same handedness as the dark photon field. In this case, outgoing radiation is peaked near the poles of the circularly polarized soliton.
\end{itemize}

Since parametric resonance can occur even for an isolated vector soliton in vacuum, it provides a channel for vector solitons to decay.  
We find that electromagnetic radiation by parametric resonance exhausts the soliton's energy very quickly as compared to the age of the Universe. 
Echoing earlier studies of scalar solitons~\cite{Hertzberg:2020dbk}, we conclude that isolated vector solitons in the Universe today must have a sufficiently small mass so as to avoid activating parametric resonance~\eqref{eq:Mc}: $M < M_c$ with $M_c \simeq (3 \times 10^{21} \ \mathrm{kg}) (m / 10^{-6} \ \mathrm{eV})^{-1} (g / 10^{-10} \ \mathrm{GeV}^{-1})^{-2/3}$. 
    
The work presented in this article furthers the effort to model vector solitons in dark matter halos, and assess their electromagnetic radiation as a potential channel for discovery.  Although isolated solitons would not be emitting, electromagnetic emission may occur when solitons collide and merge~\cite{Hertzberg:2020dbk,Du:2023jxh}.
The frequency of this radiation is controlled by the dark photon mass, $\nu \sim (200 \ \mathrm{MHz}) (m / 10^{-6} \ \mathrm{eV})$, falling into the radio band for typical masses.  
The strongest potential signal would correspond to an $O(1)$ fraction of the critical mass $M_c$ being liberated in a sudden burst of electromagnetic radiation that lasts for a time $\tau \sim \mu_\mathrm{max}^{-1}$ set by the resonance growth rate.  
Such a strong signal would easily exceed the sensitivity of typical radio telescopes, even for a cosmologically-distant source.  
To derive robust predictions for the expected signal, a more careful study of the complex dynamics of vector soliton collisions is warranted, including the effects of backreaction from electromagnetic radiation (similar to \cite{Amin:2022pzv}, but for dilute vector solitons).  
Of particular interest is the polarization of the emitted radiation, which carries information about the nature of the source, and could help to observationally distinguish vector soliton mergers from other objects.  While current radio telescopes routinely characterize the polarization of incoming radio waves \cite{2017isra.book.....T}, we have not attempted to assess the feasibility of measuring the polarization signals from solitons in a realistic setting in this paper.

It is worth reemphasizing that polarization patterns of the electromagnetic radiation depend on the nature of the interaction (the particulars of the dimension-6 operator), as well as the polarization state of dark photon field of the soliton. By contrast, radiation from scalar solitons do not show any preference for polarization of the outgoing photons. The rich structure seen in the results is a direct consequence of the assumed spin-1 nature of the dark photon field. Motivated by our results, and taking an optimistic view, if such radiation is detected, it is a potential probe of the underlying spin of the dark matter field that makes up the solitons.

Although we have focused on dark photon dark matter forming vector solitons, our analysis can be extended to other field configurations as well.  
We briefly discuss the resonance phenomenon and its implications for fuzzy dark photon dark matter in \aref{app:fuzzy}.

\begin{acknowledgments}
We are grateful to Nathaniel Craig and Ryan Plestid for guidance in discussions of the UV embedding.  
A.J.L. and E.D.S. are supported in part by the National Science Foundation under Award No.~PHY-2114024. 
M.A.A. is partially supported by a DOE grant DE-SC0021619. 
Portions of this work were conducted in the Department of Physics at the University of Jyv\"askyl\"a, Finland, and supported in part by the Academy of Finland grant 318319. 
\end{acknowledgments}

\appendix
\label{Appendix}

\section{Details of the Floquet analysis}
\label{app:details}

Following the general framework to compute Floquet solutions detailed in Sec.\,\ref{sub:floquet}, the reader can reproduce the maximum Floquet exponents
(real part) listed in eqs.\,(\ref{eq:muk_linear}) and (\ref{eq:muk_circular}) for operators $\mathcal{O}_1$ through $\mathcal{O}_5$. 
As an example, here we give a detailed derivation for operator $\mathcal{O}_3$.

\subsection{Homogeneous and linearly-polarized dark photon field for $\mathcal{O}_3$}
\label{apphomo:linear}

We consider a homogeneous and linearly-polarized dark photon field in the $\hat{z}$-direction as shown eq.\,(\ref{eq:linearsoliton}).
We use the reduced system \eqref{eq:tOPQ_equation}; however, instead of eliminating $A_3$ we eliminate $A_1$ from the system using the Coulomb gauge condition.  
Correspondingly, the $\tilde{\mathbb{O}}_{ij}=\mathbb{O}_{ij}-\mathbb{O}_{i1}k_j/k_1$ where $i$ and $j$ both take values of 2 or 3 (similarly for $\tilde{\mathbb{P}}$ and $\tilde{\mathbb{Q}}$). 
For the case under consideration, $A_3$ decouples from $A_2$, and satisfies 
\begin{align}
    &\left[ 1 -g^2\bar{X}^2(1 + \cos(2 m t))\sin^2\theta \right] \ddot{A}_3+
     \left[ 2 g^2 m \bar{X}^2\sin(2mt)\sin^2\theta \right] \dot{A}_3+
     \left[ k^2 - g^2k^2\bar{X}^2 (1 + \cos(2 m t)) \right] A_3=0
     \;,
     \label{eq:A3linear03}
\end{align}
where we used $k_3=k\cos\theta$.
We solve this equation in the small amplitude regime performing an harmonic expansion of the modes as
\begin{equation}
    A_3(t) = \sum_{l=-\infty}^{\infty} \tilde{A}_{3,l}(t) \, e^{i l m t} 
    \;, 
\label{eq:A3harmonic}
\end{equation}
where $\tilde{A}_{3}(t)$ is a slowly varying function so that $\ddot{\tilde{A}}_{3}(t)\approx 0$. 
There exists a spectrum of narrow resonant bands, which are equally spaced at $k^2\approx n^2m^2$ for $n=1,2,3, \cdots$. 
We replace \eref{eq:A3harmonic} into \eref{eq:A3linear03} and express all
cosine and sine factors in their exponential form. 
We collect all terms proportional to $e^{i l m t}$, $e^{i(l + 2)  m t}$, and $e^{i (l - 2 ) m t}$, and change the variable of summation so that they all take the form $e^{i l m t}$.  
We integrate over time from $t = 0$ to $2\pi/m$. 
The resultant equation is evaluated at $l = \pm 1$, since the first instability band dominates the resonance.  
Dropping $\tilde{A}_{3,\pm 3}$, the resultant system of differential equations to be solved is given by 
\begin{equation}
\begin{pmatrix}
    \dot{\tilde{A}}_{3,+} \\
    \dot{\tilde{A}}_{3,-}
\end{pmatrix}
    = 
    \begin{pmatrix}
         \tilde{\mathbb{M}}_{11}  &\tilde{\mathbb{M}}_{12}\\
         -\tilde{\mathbb{M}}_{12}  &-\tilde{\mathbb{M}}_{11}
    \end{pmatrix} 
    \begin{pmatrix}
         \tilde{A}_{3,+}\\
         \tilde{A}_{3,-}
    \end{pmatrix}
    \;,
\end{equation}
with
\begin{align}
&\tilde{\mathbb{M}}_{11}=\left(  2 i m - 2 i g^2 m \bar{X}^2 \sin^2\theta \right)^{-1}\left( -k^2 + m^2  + g^2k^2\bar{X}^2  - g^2m^2\bar{X}^2\sin^2\theta\right)\,,\\
&\tilde{\mathbb{M}}_{12}=\left(  2 i m - 2 i g^2 m \bar{X}^2 \sin^2\theta \right)^{-1}\left( \frac{1}{2}g^2 k^2 \bar{X}^2 + \frac{1}{2} g^2 m^2 \bar{X}^2 \sin^2\theta  \right)\,.
\end{align}
The two Floquet exponents (associated with the $A_3$ polarization mode function) are the complex eigenvalues of the $\tilde{\mathbb{M}}$ matrix.  
The eigenvalue with the larger real part is 
\begin{align}
    \mu_{\kvec,\text{max}} = &\dfrac{  \sqrt{ \left[ (2(k^2-m^2)-g^2\Xbar^2(3k^2-m^2\sin^2\theta)       \right]\left[ -2(k^2-m^2)+g^2\Xbar^2(k^2-3m^2\sin^2\theta) \right] }}{4m\left[1-g^2\Xbar^2\sin^2\theta\right]}\;.
\label{eq:mu3kO3linear}
\end{align}
The edges of the first instability band are defined by the condition $\mu_{\kvec,\mathrm{max}} = 0$.  
For a given $\theta$, using the expression above, we obtain the left edge $k_{l,{\rm edge}}$, the right edge $k_{r,{\rm edge}}$, the central wavenumber $k_0$, and the bandwidth $\Delta k$ to be 
\begin{align}
&k_{l,{\rm edge}}=m \frac{\sqrt{2-3g^2\Xbar^2\text{sin}^2\theta}}{\sqrt{2 - g^2\Xbar^2}}= m - \frac{ g^2 m \Xbar^2}{2} \left(1 - \frac{3\text{cos}^2\theta}{2} \right) + O(g^4)\,, \\
&k_{r,{\rm edge}}=m \frac{\sqrt{2-g^2\Xbar^2\text{sin}^2\theta}}{\sqrt{2 - 3g^2\Xbar^2}}= m + \frac{ g^2 m \Xbar^2}{2}\left(1+\frac{1}{2}\text{cos}^2\theta\right) + O(g^4)\,,\\
&k_0 = \frac{(k_{r,{\rm edge}}+k_{l,{\rm edge}})}{2} = m + \frac{1}{2} g^2m\Xbar^2\text{cos}^2\theta+ O(g^4)\,,\\
&\Delta k = (k_{r,{\rm edge}}-k_{l,{\rm edge}}) = g^2m\bar{X}^2\left(1-\frac{\text{cos}^2\theta}{2} \right)+ O(g^4)\,. 
\end{align}
We evaluate \eref{eq:mu3kO3linear} at $k = k_0$ finding 
\begin{equation}
    \mu_{\kvec,\mathrm{max}}(\theta) \approx \frac{ g^2 m \Xbar^2}{2}\, \left(1-\frac{\text{cos}^2\theta}{2}\right) + O(g^4) 
    \;.
    \label{eq:muA303linear}
\end{equation}
Even though $\mu_{\kvec,{\mathrm{max}}}(\theta)$ was calculated using the electromagnetic field equation of motion for $A_3$ alone, this expression matches the largest Floquet exponent among all possible electromagnetic mode functions. 

\subsection{Homogeneous and circularly-polarized dark photon field  for $\mathcal{O}_3$}
\label{Sec:O3circular}

We consider a homogeneous and circularly-polarized dark photon field on the $x-y$ plane as shown in \eref{eq:HCcirularly}. 
We use the reduced system \eqref{eq:tOPQ_equation}, and we focus on radiation that propagates along $\kvec = k \hat{\bm z}$ such that the Coulomb gauge condition imposes $A_3 = 0$.  
Working in a circular-polarization basis for the electromagnetic field, $A_L=(A_1 + iA_2)/\sqrt{2}$ and $A_R=(A_1 - iA_2)/\sqrt{2}$, the system of differential equations to be solved reads as
\begin{subequations}
\begin{align}
\ddot{A}_L + k^2 A_L - i\left[ \text{cos}(2mt)+i\text{sin}(2mt) \right] g^2 \Xbar^2 m \dot{A}_R=0\,,\label{eq:03circularAL}\\
\ddot{A}_R + k^2 A_R + i\left[ \text{cos}(2mt)-i\text{sin}(2mt) \right] g^2 \Xbar^2 m \dot{A}_L=0
    \;,
\label{eq:03circularAR}
\end{align}
\end{subequations}
We perform an harmonic expansion of the electromagnetic modes and focus on the first instability band to obtain
\begin{equation}
    \begin{pmatrix} \dot{\tilde{A}}_{L,+}\\
      \dot{\tilde{A}}_{R,+}\\
      \dot{\tilde{A}}_{L,-}\\
      \dot{\tilde{A}}_{R,-}
      \end{pmatrix} 
      = 
      \begin{pmatrix}
      \tilde{\mathbb{M}}_{11} & 0 & 0 & \tilde{\mathbb{M}}_{14}\\
      0 & \tilde{\mathbb{M}}_{22} & 0 & 0\\
      0 & 0 & -\tilde{\mathbb{M}}_{22} & 0\\
      -\tilde{\mathbb{M}}_{14} & 0 & 0 & -\tilde{\mathbb{M}}_{11} 
      \end{pmatrix} 
      \begin{pmatrix} \tilde{A}_{L,+}\\
      \tilde{A}_{R,+}\\
      \tilde{A}_{L,-}\\
      \tilde{A}_{R,-}
      \end{pmatrix}
      \label{eq:O3Circular}
    \end{equation}
where 
\begin{align}
   & \tilde{\mathbb{M}}_{11} = -\left(\frac{2}{g^2\Xbar^2}-\frac{g^2\Xbar^2}{2} \right)^{-1} \left(-\frac{i k^2}{g^2 m \Xbar^2} + \frac{i m}{g^2 \Xbar^2} - \frac{i g^2 m \Xbar^2}{2} \right)\,,\\
   & \tilde{\mathbb{M}}_{22}= -(2im)^{-1}(k^2-m^2)\,,\\
   & \tilde{\mathbb{M}}_{14} = -\left(\frac{2}{g^2\Xbar^2}-\frac{g^2\Xbar^2}{2}\right)^{-1}\left( \frac{ik^2}{2m}+\frac{im}{2}\right)\,.
\end{align}
The four Floquet exponents are the four complex eigenvalues of $\tilde{\mathbb{M}}$, and the one with largest real part is 
\begin{equation}
    \mu_{\kvec,\mathrm{max}}=\frac{\sqrt{k^4-2k^2m^2+m^4-g^2m^4\Xbar^4}}{m\sqrt{-4 + g^2\Xbar^4}}
    \qquad \text{(for $\kvec = k \, \hat{\bm z}$)}
    \;.
    \label{eq:mukO3circular}  
\end{equation}
The edges of the first instability band, its center in the $k$-space, and bandwidth read as
\begin{align}
    &k_{l,{\rm edge}} =m\sqrt{1 - g^2\Xbar^2} = m\left(1-\frac{g^2\Xbar^2}{2}\right) + O(g^4)\,,\\
    & k_{r,{\rm edge}} =m\sqrt{1+g^2\Xbar^2} = m\left(1+\frac{g^2\Xbar^2}{2}\right) + O(g^4)\,,\\
    &k_{0}=\frac{(k_{l,{\rm edge}}+k_{r,{\rm edge}})}{2}\approx m + O(g^4) \,,\\
    &\Delta k = (k_{r,{\rm edge}}-k_{l,{\rm {\rm edge}}}) \approx m g^2 \bar{X}^2 + O(g^4)\,.
\end{align}
Replacing $k_0$ into eq.\,(\ref{eq:mukO3circular}), one finds the largest
Floquet exponent among all possible wavenumbers as
\begin{equation}
    \mu_{\kvec,\mathrm{max}} \approx \frac{1}{2} g^2m\Xbar^2
    \qquad \text{(for $\kvec = k \, \hat{\bm z}$)}
\end{equation}
in complete agreement with numerical results. 

\section{Floquet analysis for spherical soliton profile}
\label{app:spherical}

In the main text we observed that the unstable modes have wavenumbers $k \sim m \gg m \, (\mu/m)^{1/2} \sim R^{-1}$ that are large compared to the inverse of the size of the soliton.  
Then if the Floquet exponent is large, $\mu_{\kvec,\mathrm{max}} \gg R^{-1}$, such that the amplification rate exceeds the escape rate for the radiation, we argued that the Floquet exponent can be calculated by treating the dark photon field as homogeneous.  
In this appendix, we relax the assumption of a large Floquet exponent, and we extend the Floquet analysis to account for the finite size of the polarized vector soliton. 

We provide the calculation for the operator $\Ocal_1$, since the equation to be analyzed is similar to that for the case of scalar solitons, which has been studied previously\,\cite{Hertzberg:2018zte}. For the other operators, we expect the general procedure describe below to work, however the structure of the equations will be more complicated to analyze numerically. We also expect the qualitative  results described here to carry over.

For an electromagnetic field $\Avec(t,\xvec)$ interacting with the vector soliton via operator $\Ocal_1$, the field's equation of motion (in Coulomb gauge $\dvec \cdot \Avec = 0$) is given by \eref{eq:OPQ_equation}: 
\begin{equation}
    \ddot{\Avec} 
    - \nabla^2 \Avec 
    +  g^2 X^2(r) \, 2\omega \sin(2 \omega t) \, \nabla \times \Avec 
    = 0
    \;.
    \label{eqO1:x}
\end{equation}
As we have done in the main text, here we drop gradients of the dark photon field when compared against gradients of the electromagnetic field, since $|\dvec A| \gg |\dvec X|$.  
Due to the inhomogeneous term with $X(r)$, it is cumbersome to work directly in $\kvec$-space, because the Fourier transform of \eref{eqO1:x} involves a convolution.  
Instead, we decompose the vector potential $\Avec(t,\xvec)$ onto a basis of vector spherical harmonics, and then eventually go to a one-dimensional Fourier space conjugate to the radial component alone.
A similar approach was employed previously in refs.~\cite{Hertzberg:2010yz,Hertzberg:2018zte} to study spherically-symmetric scalar solitons.  The corresponding equation in \rref{Hertzberg:2010yz} is $\ddot{\Avec} - \nabla^2 \Avec - g \omega \varphi(r) \sin( \omega t) \, \dvec \times \Avec = 0$. 
The results from that work can be carried over with the replacements: $-g \varphi(r) \rightarrow g^2 X^2(r)$ and $\omega \rightarrow 2\omega$. 

The vector spherical harmonic decomposition of the vector potential reads as 
\begin{equation}
    \Avec(t,\xvec) = \sum_{\l=0}^\infty \sum_{\m=-\l}^\l \Bigl[
    A_{\l\m}^{(Y)}(t,r) \, \Yvec_{\l\m}(\xhat) 
   + A_{\l\m}^{(\Psi)}(t,r) \, \Psivec_{\l\m}(\xhat) 
   + A_{\l\m}^{(\Phi)}(t,r) \, \Phivec_{\l\m}(\xhat) 
   \Bigr]
   \;,
   \label{eq:Axt}
\end{equation}
where $\Yvec_{\l\m}(\xhat)$, $\Psivec_{\l\m}(\xhat)$, and $\Phivec_{\l\m}(\xhat)$, are the vector spherical harmonics, and where $r = |\xvec|$ and $\xhat = \xvec/r$.   
The Coulomb gauge condition, $\dvec \cdot \Avec = 0$, imposes $r \l (\l+1) A_{\l\m}^{(\Psi)} = \partial_r(r^2 A_{\l\m}^{(Y)})$. 
For $\l = 0$ this implies $A_{\l\m}^{(Y)} = 0$, and since $\Psivec_{00} = \Phivec_{00} = 0$, the vector potential vanishes trivially, so only $\l > 0$ contributes.  


Using the vector spherical harmonics, the equation of motion \eqref{eqO1:x} decomposes into three separate equations for the three mode functions. 
Notice that the spherical Bessel functions of the first kind $j_\l(kr)$ are eigenfunctions of the Laplace operator. 
 We discard solutions built from spherical Bessel functions of the second kind $y_\l(kr)$, which are singular at the origin.  
This observation motivates the Ansatz 
\bsa{}{
    A_{\l\m}^{(Y)}(t,r) & = \int_{0}^{\infty} \! \frac{\mathrm{d}k}{2\pi} \biggl[ \frac{\sqrt{\l(\l+1)}}{kr} \, j_\l(kr) \, w_{k\l\m}(t) \biggr] \\ 
    A_{\l\m}^{(\Phi)}(t,r) & = \int_{0}^{\infty} \! \frac{\mathrm{d}k}{2\pi} \biggl[ - \frac{i}{\sqrt{\l(\l+1)}} \, j_\l(kr) \, v_{k\l\m}(t) \biggr] 
    \;,
}
where the complex mode functions $w_{k\l\m}(t)$ and $v_{k\l\m}(t)$ are labeled by $k \in (0,\infty)$, $\l \in \{1, 2, \cdots \}$, and $m \in \{ -\l, -\l+1, \cdots, \l-1, \l\}$.  
Using this Ansatz and the Coulomb gauge condition lets us write 
\bes{
    & A_{\l\m}^{(Y)}(t,r) \, \Yvec_{\l\m}(\xhat) 
    + A_{\l\m}^{(\Psi)}(t,r) \, \Psivec_{\l\m}(\xhat) 
    + A_{\l\m}^{(\Phi)}(t,r) \, \Phivec_{\l\m}(\xhat) 
    \\ & \quad 
    = 
    \int_{0}^{\infty} \! \frac{\mathrm{d}k}{2\pi} \biggl[ v_{k\l\m}(t) \, \Mvec_{k\l\m}(\xvec) 
    - w_{k\l\m}(t) \, \Nvec_{k\l\m}(\xvec) \biggr] 
}
where we've defined the vector spherical wavefunctions 
\bsa{}{
    \Mvec_{k\l\m}(\xvec) 
    & = 
    - \frac{i}{\sqrt{\l(\l+1)}} j_\l(kr) \, \Phivec_{\l\m}(\xhat) 
    \\ 
    \Nvec_{k\l\m}(\xvec) 
    & = 
    - \frac{\sqrt{\l(\l+1)}}{kr} \, j_\l(kr) \, \Yvec_{\l\m}(\xhat) 
    \\ & \qquad 
    - \biggl( 
    \frac{1}{kr} \, \sqrt{\frac{\l+1}{\l}} \, j_\l(kr) 
    - \frac{1}{\sqrt{\l(\l+1)}} \, j_{\l+1}(kr) 
    \biggr) \, \Psivec_{\l\m}(\xhat) 
    \nonumber 
    \;.
}
The vector spherical wavefunctions have the following properties:
\ba{
    \dvec \times \Mvec_{k\l\m} = - i k \Nvec 
    \ , \qquad 
    \dvec \times \Nvec_{k\l\m} = + i k \Mvec 
    \ , \quad \text{and} \quad 
    \dvec \cdot \Mvec_{k\l\m} = \dvec \cdot \Nvec_{k\l\m} = 0 
    \;,
}
and they are eigenfunctions of the Laplace operator: $\nabla^2 \Mvec_{k\l\m} = k^2 \Mvec_{k\l\m}$ and $\nabla^2 \Nvec_{k\l\m} = k^2 \Nvec_{k\l\m}$. 

\begin{figure}[t]
\centering
      \includegraphics[scale=0.37]{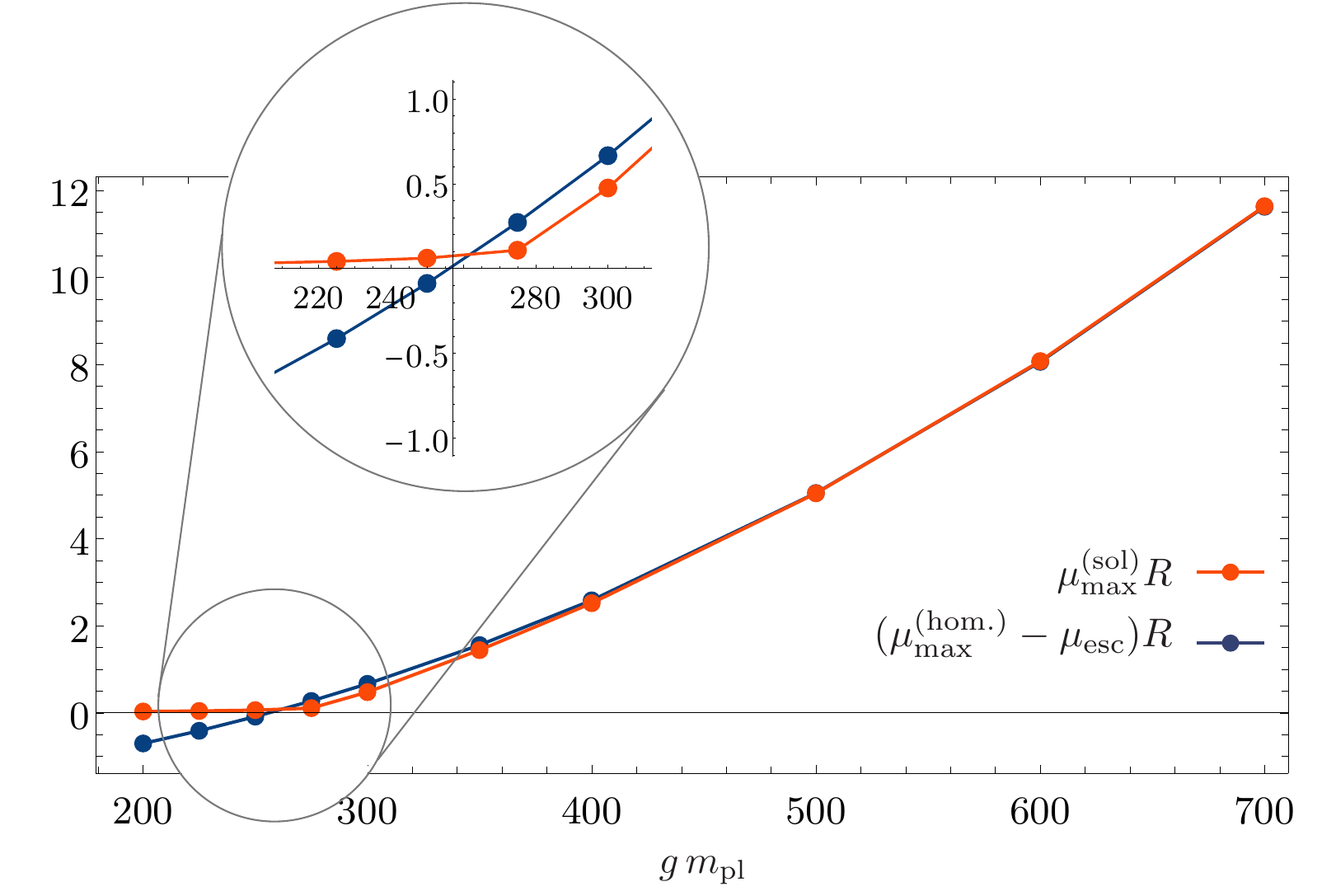}
\caption{\label{ProofRes}
The Floquet exponent with the largest real part, $\mu^{(\rm sol.)}_\mathrm{max}$, is calculated for a linearly-polarized vector soliton with $N = \mpl^2/m^2$ and coupling $g$ to electromagnetism.  The red curve shows the result of a numerical Floquet analysis applied to \eref{eq:v_eqn_Fourier}, and the blue curve shows an analytic approximation \eqref{eq:muk_approx}. We use $\mu_{\rm esc}\approx 2/R$ to match the numerical solution at large coupling.  The numerical Floquet analysis of the inhomogeneous soliton (red) confirms that the resonance starts shutting off when the escape rate becomes comparable to the homogeneous Floquet exponent (ie. when the blue curve crosses zero). In the above figure, this happens around $g\mpl\sim 250$.}
\end{figure}

The equations of motion are reduced to 
\bsa{eq:w_v_eqns}{
    & \int_{0}^{\infty} \! \frac{\mathrm{d}k}{2\pi} \biggl[ \ddot{w}_{k\l\m}(t) + k^2 \, w_{k\l\m}(t) + 2 i g^2 X^2(r) \, \omega k \, \sin(2 \omega t) \, v_{k\l\m}(t) \biggr] \times \biggl[ \frac{ \sqrt{l(l+1)}\, j_{\l}(kr)}{(kr)} \biggr] = 0\,, \\ 
    & \int_{0}^{\infty} \! \frac{\mathrm{d}k}{2\pi} \biggl[\ddot{v}_{k\l\m}(t) + k^2 \, v_{k\l\m}(t) - 2 i g^2 X^2(r) \, \omega k \, \sin(2 \omega t) \, w_{k\l\m}(t) \biggr] \times \biggl[ - i \frac{j_{\l}(kr)}{\sqrt{\l(\l+1)}} \biggr] = 0 
    \;.
}
To isolate the equation for the modes labeled with $k$, we multiply the first equation by $r^3 j_\ell(k^\prime r)$ and the second equation by $r^2 j_\ell(k^\prime r)$.  
Then integrating over $r$ and using the identity 
\ba{
    \int_0^\infty \! \mathrm{d}r \, r^2 \, j_\l(kr) j_\l(k^\prime r) = \frac{\pi}{2k^2} \, \delta(k - k^\prime) 
}
leads to 
\bsa{}{
    & \ddot{w}_{k\l\m}(t) + k^2 \, w_{k\l\m}(t) + \frac{2 k^2}{\pi} 2 i g^2 \omega\sin(2 \omega t)\int_0^\infty \! \mathrm{d}r \int_{0}^{\infty} \! \mathrm{d}k^\prime X^2(r) \, v_{k^\prime\l\m}(t) \, r^2 j_\l(k^\prime r) j_\l(k r) = 0 \\ 
    & \ddot{v}_{k\l\m}(t) + k^2 \, v_{k\l\m}(t) - \frac{2 k^2}{\pi} 2 i g^2 \omega\sin(2 \omega t)\int_0^\infty \! \mathrm{d}r \int_{0}^{\infty} \! \mathrm{d}k^\prime X^2(r) \, k^\prime  \, w_{k^\prime\l\m}(t) \, r^2 j_\l(k^\prime r) j_\l(k r) = 0 
    \;.
}
Note that the integrand contains a factor of $k$ in the first equation and a factor of $k^\prime$ in the second equation. 
The soliton profile $X^2(r)$ is defined for $r \geq 0$, and if we extend its domain to $r < 0$ by imposing $X^2(-r) = X^2(r)$, then it admits a Fourier transform 
\ba{
    X^2(r) 
    = \int_{-\infty}^{\infty} \! \frac{\mathrm{d}q}{2\pi} \, \widetilde{X^2}(q) \, e^{i q r} 
    = 2\int_{0}^{\infty} \! \frac{\mathrm{d}q}{2\pi} \, \widetilde{X^2}(q) \, \cos(qr) 
}
which lets us write 
\bsa{}{
    \ddot{w}_{k\l\m}(t) + k^2 \, w_{k\l\m}(t) + 2 i g^2 \, \omega k \, \sin(2\omega t) \, \mathcal{I}_v & = 0\,, \label{eq:me1}\\ 
    \ddot{v}_{k\l\m}(t) + k^2 \, v_{k\l\m}(t) - 2 i g^2 \, \omega k \, \sin(2\omega t) \, \mathcal{I}_w & = 0 \label{eq:me2}
    \;,
}
where
\ba{
    \mathcal{I}_v \equiv \frac{4 k^2}{\pi} \int_0^\infty \! \mathrm{d}r \int_{0}^{\infty} \! \mathrm{d}k^\prime \int_{0}^{\infty} \! \frac{\mathrm{d}q}{2\pi} \, \widetilde{X^2}(q) \, \cos(qr) \, v_{k^\prime\l\m}(t) \, r^2 j_\l(k^\prime r) j_\l(k r) 
    \;,
}
and $\mathcal{I}_w$ has $w_{k^\prime\l\m}$ instead of $v_{k^\prime\l\m}$, and it contains an additional factor of $k^\prime/k$ in the integrand.  In order to simplify $\mathcal{I}_v$, we use the identity\,\cite{Mehrem:2009ip} 
\begin{equation}
j_{\l}(kr)\,j_{\l}(k'r)=\frac{1}{2kk'r}\,\int_{|k'-k|}^{k+k'} \mathrm{d}k''\, \text{sin}(k''r)\,P_{\l}\left(\frac{k^2+k'^{2}-k''^{2}}{2kk'}\right)\,
\end{equation}
and $\int_0^\infty dr r \cos (qr)\sin (k''r)=-(\pi/2)\partial_{k''}\left[\delta(q+k'')+\delta(q-k'')\right]$, to obtain
\begin{equation}
\mathcal{I}_v  =-\int_0^{\infty} \frac{dk'}{2\pi} v_{k'\l\m}(t) \frac{k}{k'}   \int_{|k'-k|}^{k+k'} \mathrm{d}k''\, 
P_{\l}\left(\frac{k^2+k'^2-k''^2}{2kk'}\right)\, \frac{\partial }{\partial k''} \widetilde{X^2}(k'') \,.\label{eq:aux}
\end{equation}

The derivative of the one-dimensional Fourier transform $\widetilde{X^2}(k'')$ is peaked around $k''\sim 2\pi/R$ and has a width order $2\pi/R$ also. To ensure that this peak is not missed by the $dk''$ integration, we require $|k'-k| \lesssim 2\pi/R$. Furthermore, for resonance, we expect $k\approx \omega\approx m$, and recall that for non-relativistic solitons $mR\gg 1$. Note that this $m$ is mass of the dark photon, not the  index of spherical harmonic.

With these considerations, the argument of the Legendre polynomial is close to unity. Expanding the Legendre polynomial with its argument close to 1  (and for fixed $\l$) we obtain
\begin{equation}
P_{\l}\left(\frac{k^2+k'^2-k''^2}{2kk'}\right) = 1 - \frac{\l(\l+1)}{2} \left(\frac{k''^2 - (k-k')^2}{2kk'}\right)+ ...\,. 
\end{equation}
Since $(k''^2 - (k-k')^2)/(2kk')\sim 1/(mR)^2$, the Legendre polynomial is well approximated by one when $\l \ll mR$. In this regime, 
\begin{align}
\mathcal{I}_v
&=\int_0^{\infty} \frac{dk'}{2\pi} v_{k'\l\m}(t) \frac{k}{k'} \Bigl[ \widetilde{X^2}(k-k^\prime) - \widetilde{X^2}(k+k^\prime) \Bigr]+O[\l^2/(mR)^2]\,\\
&\approx \int_0^{\infty} \frac{dk'}{2\pi} v_{k'\l\m}(t) \widetilde{X^2}(k-k^\prime)+O[\l^2/(mR)^2].\label{eq:aux2}
\end{align}
where in the second line we used $k'\sim k\sim m$, and ignored $\widetilde{X^2}(k+k')$  using the fact that $\widetilde{X^2}(q)$ is centered around $q=0$ with a width $\sim 1/R\ll m^{-1}$. Also note that with the same approximations $\mathcal{I}_w\approx \mathcal{I}_v$ (with $v\rightarrow w$).

Then we may write (\ref{eq:me1}) and (\ref{eq:me2}) as 
\bes{\label{eq:v_eqn_Fourier}
    & \ddot{\tilde{v}}_{k\l\m}(t) + k^2 \, \tilde{v}_{k\l\m}(t) \pm 2 g^2 \omega k \, \sin(2 \omega t) \int_{0}^\infty \! \frac{\mathrm{d}k^\prime}{2\pi} \, \widetilde{X^2}(k-k^\prime) \, \tilde{v}_{k^\prime\l\m}(t) = 0 \\
    & \quad \text{and} \qquad 
    \tilde{w}_{k\l\m}(t) = \pm i \tilde{v}_{k\l\m}(t) \quad \text{with} \quad \l \ll mR 
    \;.
}
Note that $\widetilde{X^2}(k-k^\prime)$ couples modes over a $k'$ width of $\sim 1/R$ around $k\approx m$. 
These results agree with those derived in \cite{Hertzberg:2018zte} for the case of scalar solitons, under the replacement $2\omega \rightarrow \omega$ and $\widetilde{X^2}(k-k')\rightarrow \widetilde{\Phi}(k-k')$, where $\widetilde{\Phi}(k-k')$ represents the one dimensional Fourier transform of the scalar soliton profile. 
Authors in \cite{Hertzberg:2018zte} only studied the particular channel $(\l,\m)=(1,0)$, but due to the likeness between the vector and scalar soliton analysis, we can conclude that their results generically holds for any pair of spherical harmonic numbers $(\l,\m)$ so long as $\l \ll mR$. We note that when $\l\gg m R$, we have numerically verified that $\mathcal{I}_{v,w}$ decays exponentially with $\l$, and hence we ignore that regime in what follows.

The integro-differential eq.\,(\ref{eq:v_eqn_Fourier}) can be analysed using Floquet theory since the system is coupled in $k$-space, but still periodic in time. We will follow \rref{Hertzberg:2018zte} where a closely related system was analyzed. We discretize the system in $k$-space, and solve the coupled system of different $k$ modes numerically. There are two physical considerations which set the resolution and size of the grid in $k$-space. First, the width $\sim 1/R$ of $\widetilde{X^2}(k-k')$ sets the extent of the $k$-space grid, whereas the requirement of resolving the resonance band near $k\approx m$, sets the resolution of the $k$-grid ($\Delta k<g^2\bar{X}^2m$).  
   
By numerically solving the integro-differential equation, we study a linearly-polarized vector soliton with $N = \mpl^2/m^2$ and different values of the coupling $g$.  We calculate the Floquet exponent with the largest real part, $\mu_\mathrm{max}^\mathrm{(sol.)} = \max\limits_{i} \Re[\mu_{i}]$.  
Our results are summarized in \fref{ProofRes}, which shows the dependence of the Floquet exponent on the coupling $g$.  
These results show an excellent agreement with the analytic approximation in eq.\,(\ref{eq:muk_approx}), i.e. the resonance phenomenon is turned on  when the maximal Floquet exponent for the corresponding homogeneous
case starts becoming larger than the soliton light-crossing time,  $\mu_{\text{max}}^{\text{(hom.)}}\gtrsim O(1/R)$. The quantity $\mu_{\rm esc.}=O(1/R)$ can be interpreted as the escape rate for radiation leaving the soliton. When $\mu_{\text{max}}^{\text{(hom.)}}\lesssim \mu_{\text{esc}}$, radiation is leaves the system more quickly than it is being generated and the Bose enhancement required during the resonance is suppressed. The same feature was reported in \rref{Hertzberg:2018zte} for the case of scalar solitons.\footnote{In \rref{Hertzberg:2018zte}, the growth rate of photons in scalar solitons follows the analytical approximation $\mu_{\text{max}}^{\text{(hom.)}} - \mu_{\text{esc}}$, with the escape rate  defined as $\mu_{\text{esc}}=1/(2R_c)$. These authors approximate the scalar soliton profile using a sech function, where $R_c$ is a characteristic  scale length. This quantity is related to the radius of the power-law approximation, eq.\,(\ref{eq:EMR}), as $R_c \approx R/3$. In addition, the resonance calculation for the scalar   case considers one power of the  soliton profile, while that for the vector case involves  the square of the soliton profile. The ratio between the full width at half maximum of a sech function and its square is about $\sqrt{2}$. Thus, transforming the scalar escape rate which fits numerical data to the vector escape rate which fits ours, we have $1/(2R_c) \rightarrow 3\sqrt{2}/(2R)\approx 2/R$, in complete agreement with results shown in \fref{ProofRes}.} We find that $\mu_{\rm max}^{(\rm sol.)}\approx \mu_{\text{max}}^{\text{(hom.)}}-\mu_{\rm esc}$ if we fix $\mu_{\rm esc}\approx 2/R$. 

We note that the zoom-in \fref{ProofRes}  shows a slightly  disagreement between numerical results at $g\mpl <250$ which are small but non-zero (at a level larger than machine precision), and the analytical expectation that these should approach zero. Even a small non-zero Floquet exponent is relevant because of the exponential nature of the instability, and required further analysis to determine whether this discrepancy is physical or numerical. Our analysis indicates that this disagreement is a result of numerical issues. Resolving the resonance band $\Delta k\propto g^2\bar{X}^2m$, and covering the width $\sim 1/R$ of $\widetilde{X^2}(k-k')$ becomes exceptionally challenging at small $g$. We have found that for $g\mpl\lesssim 250$, the numerically evaluated Floquet exponent continues to decrease as we increase the resolution and extent of the $k$-grid, whereas for larger $g$ the values do not change. While not quite a proof, we take this as an indication that the Floquet rate approaches zero for small coupling as expected from theoretical considerations.

\section{Fuzzy dark photon dark matter}
\label{app:fuzzy}

Although our primary interest in this work has been the phenomenon of parametric resonance in polarized vector solitons, the calculations presented here can be carried over to other systems as well.  
In this appendix, we consider fuzzy dark photon dark matter, not forming solitons, and we adapt the results of our analysis to assess the implications of parametric resonance of electromagnetic radiation for this system.  

The inhomogeneous dark photon field admits a Fourier representation as 
\ba{
    \Xvec(t,\xvec) = \int \! \! \frac{\mathrm{d}^3 \kvec}{(2\pi)^3} \, \Xvec_\kvec(t) \, e^{i \kvec \cdot \xvec}
    \;,
}
where modes are labeled by a wavevector $\kvec$ with corresponding wavenumber $k = |\kvec|$ and wavelength $\lambda = 2\pi/k$.  
We are interested in systems in which the dark photons are non-relativistic, which means that the modes amplitudes $\Xvec_\kvec(t)$ only have support for modes with small wavenumbers $k \ll m$.  
As a fiducial parameter choice we take $m = 10^{-20} \ \mathrm{eV}$, corresponding to `fuzzy' dark matter, and the non-relativistic modes have $\lambda \gg 2\pi/m \simeq (0.004 \ \mathrm{pc}) (m / 10^{-20} \ \mathrm{eV})^{-1}$. 

The energy density carried by the non-relativistic dark photon field today is approximately $\rho_X(\xvec) \approx m^2 |\Xvec(\xvec)|^2 / 2$.  
Assuming that the mode amplitude is only a function of the wavenumber, $|\Xvec_\kvec| = |\Xvec_k|$, we can write 
\bes{
    \int \! \mathrm{d}^3 \xvec \, \rho_X(\xvec) 
    & \approx \int_0^\infty \! \frac{\mathrm{d} k}{k} \, \mathcal{E}_{X,k} 
    \qquad \text{with} \qquad 
    \mathcal{E}_{X,k} = \frac{1}{4\pi^2} m^2 k^3 |\Xvec_k|^2
    \;,
}
where $\mathcal{E}_{X,k}$ is the spectral energy distribution of the dark photon field today.  
We assume that $\mathcal{E}_{X,k}$ is peaked at a wavenumber $k = k_\ast$ such that $0 < k_\ast \ll m$.  
Consequently, non-relativistic modes with wavevectors satisfying $|\kvec| \approx k_\ast$ carry most of the energy.  
For instance, the production mechanism discussed in Refs.~\cite{Graham:2015rva,Ahmed:2019mjo,Ahmed:2020fhc,Kolb:2020fwh} leads to $k_\ast \sim \sqrt{m H_\mathrm{eq}}/(1+z_\mathrm{eq})$
where $z_\mathrm{eq} = 3300$ and $H_\mathrm{eq} = 10^{-28} \ \mathrm{eV}$ are the redshift and Hubble parameter at radiation-matter equality; this corresponds to a length scale today of $\lambda_\ast \simeq ( 63 \ \mathrm{kpc} ) (m / 10^{-20} \ \mathrm{eV})^{-1/2}$. 
We define $\Xbar = k_\ast^3 |\Xvec_{k_\ast}| / 2\pi^2$ to be the field amplitude of the dominant modes, and their energy density is written as $\rho_{X,\ast} \approx \mathcal{E}_{X,k_\ast} \approx \frac{1}{2} m^2 \Xbar^2 $.  

The condition $\mu_\mathrm{max}^\mathrm{(hom.)} \, \lambda_\ast \approx g^2 \Xbar^2 m \lambda_\ast / 2 \gtrsim 1$ must be satisfied in order for parametric resonance to occur; see \eref{eq:muk_approx}.  
This places a lower bound on the field amplitude $\Xbar$ that depends upon the coupling $g$, mass $m$, and the coherence length scale $\lambda_\ast$.  
Conversely, the requirement that the dark photon relic abundance does not exceed the known dark matter relic abundance imposes an upper bound on the field amplitude today $\Xbar \leq \sqrt{2 \rho_\text{\sc dm} / m^2}$.   
Taken together, these two bounds are expressed as 
\ba{\label{eq:fuzzy_bounds} 
    \bigl( 1 \times 10^{4} \ \mathrm{GeV} \bigr) \biggl( \frac{g}{10^{-10} \ \mathrm{GeV}^{-1}} \biggr)^{-1} \biggl( \frac{m}{10^{-20} \ \mathrm{eV}} \biggr)^{-1/2} \biggl( \frac{\lambda_\ast}{1 \ \mathrm{Gpc}} \biggr)^{-1/2} 
    \ll 
    \Xbar 
    \leq 
    \bigl( 4 \times 10^{5} \ \mathrm{GeV} \bigr) \biggl( \frac{m}{10^{-20} \ \mathrm{eV} } \biggr)^{-1}
    \;.
}
These inequalities emphasize why we focus on such low-mass fuzzy dark matter with $m \sim 10^{-20} \ \mathrm{eV}$.  
For larger values of $m$ (at the same $g$, $\lambda_\ast$) the upper and lower bounds become incompatible.  

If the condition for parametric resonance is satisfied, the dark photon field will decay into electromagnetic radiation.  
The time scale for this energy transfer is controlled by the maximal Floquet exponent via $\tau \approx 1 / \mu_\mathrm{max}^\mathrm{(hom.)} \approx 2/ g^2 \Xbar^2 m$, using the results in \erefs{eq:muk_linear}{eq:muk_circular}.  
For the same fiducial parameters used in the estimates above, we have the lifetime 
\ba{
    \tau \simeq \bigl( 1 \times 10^{15} \ \mathrm{s} \bigr) \biggl( \frac{g}{10^{-10} \ \mathrm{GeV}^{-1}} \biggr)^{-2} \biggl( \frac{\Xbar}{10^{5} \ \mathrm{GeV}} \biggr)^{-2} \biggl( \frac{m}{10^{-20} \ \mathrm{eV}} \biggr)^{-1} 
    \;.
}
Since the age of the universe today is $t_0 \sim 10^{17} \, \mathrm{s}$, these estimates imply that the dark photon field would have been depleted long ago by the resonant amplification of electromagnetic radiation.  
Conversely, the condition for parametric resonance to be inoperative today is written as 
\ba{\label{eq:no_PR_for_fuzzy} 
    \biggl( \frac{g}{10^{-10} \ \mathrm{GeV}^{-1}} \biggr)^{}
    < 
    \bigl( 0.025 \bigr) 
    \biggl( \frac{m}{10^{-20} \ \mathrm{eV}} \biggr)^{1/2} \biggl( \frac{\lambda_\ast}{1 \ \mathrm{Gpc}} \biggr)^{-1/2} 
    \;,
}
assuming that the dark photon makes up all of the dark matter.  
Therefore, in order to have a viable model of fuzzy dark photon dark matter coupled to electromagnetism through the dimension-6 operators that we consider, the parameters must be such that the parametric resonance instability is inoperative today, implying an upper limit on the coupling $g$. 

If the parameters are chosen such that the parametric resonance instability is inoperative today, it is interesting to ask whether parametric resonance may have taken place in the early universe.  
Specifically, we are interested in the time dependence of $\mu_\mathrm{max}^\mathrm{(hom.)}(t) \lambda_\ast(t)$ and how it compares to $1$. 
In an Friedman-Robertson-Walker spacetime with scale factor $a(t)$ at time $t$, we can write the time dependence as $\mu_\mathrm{max}^\mathrm{(hom.)}(t) \lambda_\ast(t) \propto a(t)^r \Xbar(t)^2 \lambda_\ast(t)$ where the additional factors of $a(t)^r$ arise from the metrics and inverse metrics appearing in the operators $\Ocal_1$ through $\Ocal_5$; for example, $X \cdot X = g^{\mu\nu}(t) X_\mu X_\nu \approx a(t)^{-2} \Xbar(t)^2$.  
The field's coherence length grows no more quickly than $\lambda_\ast(t) \propto a^2(t)$ (tracking the causal horizon during radiation domination); we can write $\lambda_\ast(t) \propto a(t)^s$.
Similarly, the field amplitude (for non-relativistic modes inside the horizon) oscillates under a decreasing envelope $\Xbar(t) \propto a(t)^{-1/2}$~\cite{Graham:2015rva}. 
Putting together these factors gives $\mu_\mathrm{max}^\mathrm{(hom.)}(t) \lambda_\ast(t) \propto a(t)^{r + s - 1}$. 
Since $a(t)$ is a growing function of time, if the condition for parametric resonance is not satisfied today, and if $r + s - 1 \geq 0$ then parametric resonance was never operative (on cosmological scales) throughout the cosmic history.  
Conversely, if $r + s - 1 < 0$ then parametric resonance may have taken place in the early universe.  
The associated electromagnetic energy injection to the primordial plasma may have had an observable impact on the abundances of light elements, produced at big bang nucleosynthesis (BBN)~\cite{Fields:2019pfx}, or on spectral distortions of the cosmic microwave background (CMB) radiation \cite{Chluba:2011hw}.  

 \bibliographystyle{JHEP.bst}
\bibliography{v1_ALS23a.bib}

\end{document}